\documentclass[twocolumn,superscriptaddress,nobibnotes,nofootinbib,floatfix,aps,prl,citeautoscript,reprint,longbibliography]{revtex4-2}

\usepackage[utf8]{inputenx}
\usepackage[T1]{fontenc}
\usepackage{graphicx}
\usepackage{epsfig}
\usepackage{subfigure}
\usepackage{color}
\usepackage{latexsym}
\usepackage{amsmath,bm,multirow}
\usepackage{amsfonts}
\usepackage{dcolumn}           
\usepackage{natbib}
\usepackage{physics}

\usepackage[draft]{changes} % final OR draft
\usepackage[normalem]{ulem}

\newcommand{\nw}{Department of Materials Science and Engineering, Northwestern University, Evanston, IL 60208, USA}

\usepackage[draft]{changes} % final OR draft
\usepackage[normalem]{ulem}

\usepackage{soul}
\usepackage{lipsum}
\makeatletter
\def\blfootnote{\xdef\@thefnmark{}\@footnotetext}
\makeatother

\begin{document}

\title{Scale-invariant Machine-learning Model Accelerates the Discovery  of  Quaternary Chalcogenides with  Ultralow Lattice Thermal Conductivity}

\author{Koushik Pal}
\email{koushik.pal.physics@gmail.com}
\affiliation{\nw}

\author{Cheol Woo Park}
\blfootnote{K.P. and C.W.P. contributed  equally.}
\email{cheolpark2016@u.northwestern.edu}
\affiliation{\nw}

\author{Yi Xia}
\affiliation{\nw}

\author{Jiahong Shen}
\affiliation{\nw}

\author{Chris Wolverton}
\email{c-wolverton@northwestern.edu}
\affiliation{\nw}

%\date{\today}

%\date{\today}

\begin{abstract} 
Intrinsically low lattice thermal conductivity  ($\kappa_l$)  is  a  desired requirement  in  many  crystalline solids such as thermal barrier coatings and thermoelectrics. Here, we design an advanced machine-learning (ML) model based on crystal graph convolutional neural network that is insensitive to volumes (i.e., scale) of the input crystal structures to discover novel quaternary chalcogenides, AMM'Q$_3$ (A/M/M'=alkali, alkaline-earth, post-transition metals, lanthanides, Q=chalcogens). Upon screening the thermodynamic stability of $\sim$ 1 million compounds using the ML model iteratively and performing density functional theory (DFT) calculations for a small fraction of compounds, we discover 99 compounds that are validated to be stable in DFT. Taking several DFT-stable compounds, we calculate their $\kappa_l$ using phonon-Boltzmann transport equation, which reveals ultralow-$\kappa_l$ ($<$ 2 Wm$^{-1}$K$^{-1}$ at room-temperature) due to their soft elasticity and strong phonon anharmonicity.  Our work demonstrates the high-efficiency of scale-invariant  ML  model  in predicting novel compounds and presents experimental research opportunities with these new compounds.
\end{abstract}
%%%%%%%%%%%%%%%%%%%%%%%%%%%%%%%%%%%%%%%%%%%%%%%%%%%%%%%%%%%%%%%%%%%

\maketitle

\section{Introduction} 
 
The study of heat transport phenomena in materials  is imperative  to find and deploy suitable materials in various thermal-energy  management platforms  such as thermal barrier coatings \cite{wu2002low}, waste-heat  recovery  devices \cite{bell2008cooling} and modern  high-performance computing architectures which  require rapid heat dissipation  \cite{li2018high}.   A sustained research  effort in  this direction has focused  on finding materials with  extreme thermal transport  properties \cite{lindsay2013first,li2018high,tian2018unusual,samanta2020intrinsically,mukhopadhyay2018two,xia2020particlelike}.  Among them,  semiconducting materials with very low lattice thermal conductivity ($\kappa_l$)  are particularly interesting  as they find  applications in  thermoelectrics (TEs) which can convert heat into electrical energy \cite{biswas2012high,zhao2014ultralow,slade2020contrasting}.  The  TE conversion efficiency of  materials, quantified by the figure of merit, $ZT = \frac{S^2\sigma}{(\kappa_l+\kappa_e)}T$  can be  enhanced by reducing their $\kappa_l$. In the  preceding equation $S$, $\sigma$, and $\kappa_e$ are the Seebeck coefficient, electrical conductivity, and  electrical contribution to the total thermal  conductivity ($\kappa = \kappa_l + \kappa_e$), respectively.   Hence, novel compounds with intrinsically  low $\kappa_l$ are highly sought  after for fundamental research that would help in the design and discovery of efficient materials suitable for device  applications.

The discovery of novel compounds  is an important  yet challenging task in materials  science.  Traditionally, trial and error methods have been employed in the laboratory  to synthesize new compounds. However,  accurate quantum mechanical  methods such as density functional theory (DFT) have  proven to be extremely beneficial in the discovery of new materials and estimation of their  properties, by reducing  the target composition space considered for exploratory synthesis in the laboratory. In the last few years, modern computational approaches such as  high-throughput (HT) DFT calculations based on prototype decoration  has accelerated  the discovery of new compounds  \cite{curtarolo2013high,  jain2016computational, saal2013materials, jain2013commentary, curtarolo2012aflowlib}. These approaches take advantage of the computed energetics of  large number of  materials available in diverse materials databases such as the Open Quantum Materials Database (OQMD) \cite{saal2013materials, kirklin2015open}, Materials Project \cite{ jain2013commentary},  Aflowlib \cite{curtarolo2012aflowlib},  which enable us to perform a highly accurate phase stability analysis of a novel composition taking into account all of its  competing  phases. 

In the HT-DFT method,  the initial crystal structures are generated by decorating  prototype crystal structures with elements from the periodic table. To keep the number of calculations of  generated  compounds computationally tractable, we often  use certain rules during prototype decoration,  which are typically derived from examining the already known  compounds in that family.  For example,   (a) chemically similar elements  are substituted at  each crystallographically inequivalent site in the prototype structures and (b) only those  compositions which balance the  valence charges of the elements   are generated.   While these considerations often lead to  a high  success rate in predicting stable  compounds,  it leaves  out a large number of other ``unsuspected" compounds  that would be generated if we  substitute all elements  from the periodic table in any sites  of the prototype structures in all possible combinations. Therefore, we suffer the risk of  missing out perhaps many hitherto unknown stable  compounds that could exhibit exciting physical and chemical properties.

Machine learning (ML) offers a computationally feasible solution to this problem. Using ML methods, the entire phase space of multinary compositions can be quickly  screened for possible stable and metastable compounds even before doing any expensive DFT calculations.  In recent years, ML methods  have proven to be an invaluable tool in discovering novel compounds in  multi-component composition spaces \cite{rupp2015machine,ward2016general,faber2016machine, faber2015crystal, hautier2010finding, tabor2018accelerating, meredig2014combinatorial, balachandran2018predictions, ramprasad2017machine, ward2017atomistic}.  In one successful example, Ren et al. \cite{ren2018accelerated} predicted new stable bulk metallic glasses using an ML model and  were able to experimentally synthesize them. Another interesting  example  of  ML-guided materials discovery is demonstrated  in the work by   Kim et al. \cite{kim2018machine}  where they predicted  several new stable quaternary  Heuslers (QH).  In their work, Kim et al. generated  nearly 3.3 million quaternary compositions in the  Heusler structure by substituting 73 metallic elements from the periodic table  at the three inequivalent crystallographic sites in every possible way. An ML  model was constructed   which was  trained on the computed energetics  of compounds available in  the OQMD   to assess the phase stability  of  those 3.3 million compositions.  The ML-predicted stable  compounds  were  then validated by performing DFT   calculations which  gave rise to 55 new DFT-stable  compounds  that  were  missing from the earlier HT-DFT \cite{gautier2015prediction,he2018designing} and ML  \cite{balachandran2018predictions} works in the same Heusler family as those previous works only explored a smaller set of  compositions restricted  by  the electron-counting  and  charge-neutrality  conditions.

In this work,   we use an ML method to  explore the vast  phase space spanned by a family of experimentally known quaternary chalcogenides (AMM'Q$_3$) \cite{koscielski2012structural,strobel2006three,ruseikina2019synthesis,maier2016crystal,ruseikina2017trends,ruseikina2018refined,sikerina2007crystal,prakash2015syntheses} that possess diverse structure types and chemistry. Some of these compounds are  shown to exhibit very low lattice thermal conductivity \cite{pal2019intrinsically, hao2019design},  promising thermeoelectric performance \cite{pal2019high, pal2019unraveling} and high photovoltaic  efficiency \cite{fabini2019candidate}.  In our previous HT search \cite{pal2021accelerated} for new materials in this crystal family, we  discovered a large number (628) of  stable compounds. However, our previous search \cite{pal2021accelerated}  explored only a small set of possible compositions since  we generated the initial crystal  structures of the  compounds  following a set of rules that were derived by examining all experimentally known AMM'Q$_3$ compounds. They are: (a) all elements in the  AMM'Q$_3$ compounds  are in  their most common oxidation states which satisfy  the valence-charge-neutrality condition,  (b) the A-site is occupied by alkali, alkaline-earth, or post-transition metals with the only exception  of Eu which is a lanthanide, (c) the M-site is occupied by transition   metals,  d) the M' site is occupied by  transition metals or lanthanides, (e) Q  site is always occupied by  S, Se, or Te, and (f) no  compounds contain more than one alkali and alkaline earth metals or a combination of them.  Adhering to these criteria, we generated 4659 unique charge-balanced compositions and performed HT-DFT calculations  considering all structural prototypes that are known in this family of compounds,  leading to the discovery of 628 stable and 852 low-energy metastable compounds in our previous work \cite{pal2021accelerated}.

Thus, the chemical trends that can be observed from the experimentally known AMM'Q$_3$ compounds served as a  useful guide to  discover  previously unknown stable compounds in this family of materials. However, in this study, we are interested in identifying new stable  compositions in the AMM'Q$_3$ prototypes that do not necessarily follow those  chemical criteria and therefore, were overlooked in the previous  HT-DFT search \cite{pal2021accelerated}. For an exhaustive search of the phase space, in this work, we generate the initial quaternary compositions that do not necessarily follow the previous rules. Here,  66 metallic elements are considered for substitutions at the A, M, and M’ cation sites in every possible way while generating the initial structures. Keeping the Q atoms fixed to  three chalcogens (S, Se, and Te), we generate  a total number of  823,680  ( = 3 $\times$ $^{66}$P$_3$) initial compositions of these quaternary chalcogenides which is the target search space in this work.

To this end,  we  develop an advanced ML framework  based  on the recently proposed iCGCNN framework \cite{park2020developing}, a variant model of the crystal graph convolutional neural network  (CGCNN)  \cite{xie2018crystal}. In the iCGCNN framework, crystal structures are  represented  as crystal graphs which are then used  as input for graph neural networks to predict material properties of interest.  Although iCGCNN has been shown to  exhibit high accuracy in predicting the formation energy, a property directly  relevant to the thermodynamic stability of a  material, here, we show that iCGCNN exhibits  peak performance when the  input  crystal  structures are fully relaxed in terms of volume,   stress, and ionic positions. This  dependency can limit the effectiveness of the ML models in situations where   the relaxed crystal structures are  unavailable. This situation can arise when  prior HT-DFT  data for a class of materials do not exist. For our ML  model used in this study, we designed it such that the formation energy   predictions are invariant to the volume  (i.e., scale)  of the input crystal  structures to account for the fact that unrelaxed  crystal structures can have arbitrary  volumes. We show that our ML  model outperforms the iCGCNN model by 25\%  in predicting the formation energies  of materials when provided with the unrelaxed crystal structures as input. 
 
With the iterative use of the ML model  on the 823,680 newly generated  compounds combined with successive filtering and DFT calculations,  we  discover hitherto unknown 99 DFT-stable  and 362 low-energy DFT-metastable compounds that are potentially synthesizable in the laboratory. Of the newly discovered 99 stable compounds, we  randomly chose 15  compounds that are semiconducting and non-magnetic to examine their thermal transport properties  by solving the  phonon Boltzmann transport equation (PBTE) in a first-principles framework. The newly discovered compounds in this work are validated by DFT to possess ultralow $\kappa_l$, and have ``unsuspected" combination of elements, and hence are different from those predicted in a previous HT-DFT work \cite{pal2021accelerated}.

\begin{figure*}
\centering
\includegraphics[width=1.0\textwidth]{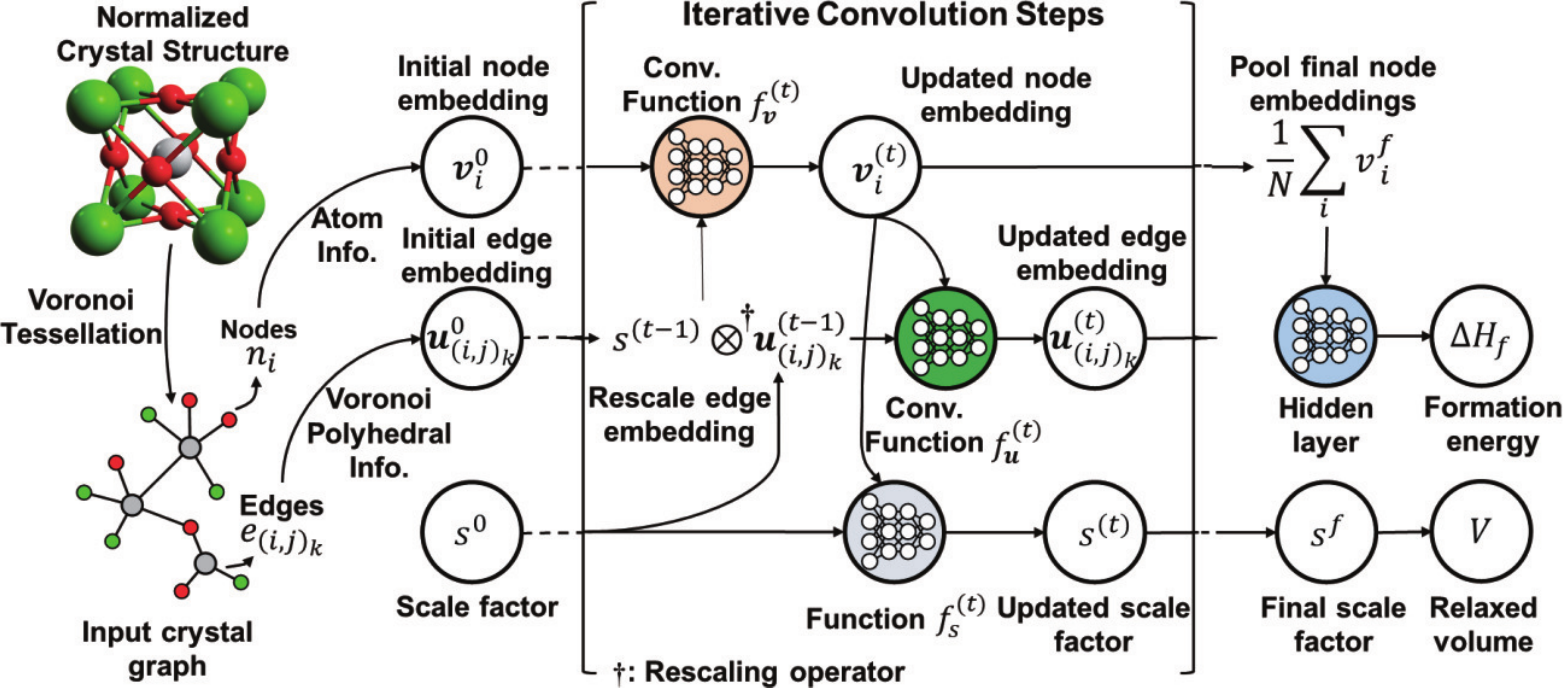}
\caption{\textbf{Schematic design of the machine-learning model}. Illustration of the iCGCNN-based multi-objective ML framework that predicts both formation energy and relaxed volume of a crystal. The crystal graph is constructed from the Voronoi tessellated crystal structure that has been normalized to have a minimum interatomic distance of 1. Additional to the node and edge embeddings, $\boldsymbol{v}_i$ and ${\boldsymbol{u}_{(i,j)}}_k$, that encode the atom and bond information, the crystal graph is associated with scale factor s that represents the minimum interatomic distance of the crystal. In each iterative convolution step, $s$ is updated as a function of $\boldsymbol{v}_i$ such that, at the end of all iterations, the final value $s^f$ matches the minimum interatomic distance that would be measured in the crystal structure that has been relaxed with respect to volume. $\boldsymbol{v}_i$ and ${\boldsymbol{u}_{(i,j)}}_k$ are also iteratively updated to better represent the local chemical environment of the crystal. Formation energy is predicted from the final node embedding $\boldsymbol{v}_i^f$ and relaxed volume of the crystal is calculated by multiplying the cube of $s^f$ with the volume of the normalized crystal structure. }
\end{figure*}

\section{Results and discussion}

\subsection{Scale-invariant machine learning model}

We design our ML model based  upon iCGCNN, a variant model of the crystal graph  convolutional  neural network (CGCNN) \cite{xie2018crystal, park2020developing}. In iCGCNN,  the unit cell of a compound is represented as a crystal graph $\mathcal{G}$ = ($\mathcal{N, E}$), where node  n$_i$ $    \in  \mathcal{N}$ represents constituent atom $i$ and edge  ${e_{(i,j)}}_k$ $\in$ $\mathcal{E}$ represents the bond between atom $i$ and neighboring atom $j$. Atoms are considered  neighbors if they share a face in the Voronoi tessellated crystal structure. To account for the periodicity of the crystal,  multiple edges can exist between 
neighboring nodes as indexed by $k$. Node n$_i$ is then embedded with vector $\boldsymbol{v}_i$ that encodes the elemental properties of atom $i$, where embedding is defined as the mapping of a discrete object  to a vector of real numbers. Edge ${e_{(i,j)}}_k$ $\in$ $\mathcal{E}$ is also embedded with vector ${\boldsymbol{u}_{(i,j)}}_k$ $\in$ $\mathcal{E}$  that encodes the structural  information of the polyhedra formed by neighboring atoms $i$ and $j$, and their shared Voronoi face. Structural
information encoded in ${\boldsymbol{u}_{(i,j)}}_k$ $\in$ $\mathcal{E}$  include the interatomic distance between atoms $i$ and $j$,  solid angle of atom $i$ with  respect to the shared Voronoi face, area of the shared Voronoi face, and volume of the polyhedra. This crystal graph is then used as direct input to a graph neural network where the node and edge embeddings are iteratively updated according to pre-defined convolution functions. The convolution functions for the node ($f_{\boldsymbol{v}}^t$) and edge ($f_{\boldsymbol{u}}^t$) embeddings at the $t$-th iteration are given by:

%\lipsum[1]
\begin{widetext}
\begin{align}
\begin{split}
f_{\boldsymbol{v}}^t: \boldsymbol{v}_i^{(t+1)}=\boldsymbol{v}_i^{(t)}+ \sum_{(j,k)} \sigma ({{\boldsymbol{z}^{(t)}_{(i,j)}}_k} \boldsymbol{W}{_1}^{(t)}+\boldsymbol{b}{_1}^{(t)} ) \odot g{{(\boldsymbol{z}^{(t)}}_{(i,j)}}_k\boldsymbol{W}_{2}^{(t)}  + \boldsymbol{b}_2^{(t)})  + \\ \sum_{(j,l,k,k')}{\sigma({\boldsymbol{z}^{(t)}}_{(i,j,l)}}_{(k,k')} \boldsymbol{W}_1^{(t)}+\boldsymbol{b}_1^{(t)} ) \odot g{{(\boldsymbol{z}^{(t)}_{(i,j,l)}}_{(k,k')} } \boldsymbol{W}_2^{(t)}+  \boldsymbol{b}_2^{(t) } )
\end{split}
\end{align}
%\end{widetext}
%\lipsum[1]

%\begin{widetext}
\begin{align}
\begin{split}
f_{\boldsymbol{u}}^t: \boldsymbol{u}_i^{(t+1)}=\boldsymbol{u}_i^{(t)}+ \sum_{(j,k)} \sigma ({{\boldsymbol{z}^{(t)}_{(i,j)}}_k} \boldsymbol{W}{_1}^{(t)}+\boldsymbol{b}{_1}^{(t)}) \odot g{{(\boldsymbol{z}^{(t)}}_{(i,j)}}_k\boldsymbol{W}_{2}^{(t)}  + \boldsymbol{b}_2^{(t)}) \\ + \sum_{(j,l,k,k')}{\sigma({\boldsymbol{z}^{(t)}}_{(i,j,l)}}_{(k,k')} \boldsymbol{W}_1^{(t)}+\boldsymbol{b}_1^{(t)} ) \odot g{{(\boldsymbol{z}^{(t)}_{(i,j,l)}}_{(k,k')} } \boldsymbol{W}_2^{(t)}+  \boldsymbol{b}_2^{(t) }).
\end{split}
\end{align}
\end{widetext}

In the above equations, $\odot$ represents an element-wise matrix multiplication while $\sigma$ and $g$
 represent a sigmoid function and a nonlinear activation function, respectively. $\boldsymbol{W}^{(t)}$ and $\boldsymbol{b}^{(t)}$ represent the weight and bias matrices respectively for the $t$-th convolution step.  ${{\boldsymbol{z}^{(t)}_{(i,j)}}_k}=\boldsymbol{v}_i^{(t)} \oplus \boldsymbol{v}_j^{(t)} \oplus {\boldsymbol{u}}_{{(i,j)}_k}^{(t)}$ is the concatenation of the node and edge vectors and captures the two-body correlation of atoms $i$ and $j$. 
Likewise, the node vectors and edge vectors that connect atoms $i$, $j$, and $l$ are concatenated to form
${\boldsymbol{z}'^{(t)}_{(i,j,l)}}_{(k,k')}=\boldsymbol{v}_i^{(t)} \oplus \boldsymbol{v}_j^{(t)} \oplus \boldsymbol{v}_l^{(t)} \oplus  {\boldsymbol{u}^{(t)}_{(i,j)}}_k \oplus {{\boldsymbol{u}^{(t)}_{(i,l)}}_{k'}}$ to capture the three-body correlations of the atoms.  After each iteration of convolution, the node and edge embeddings are updated to better represent the local chemical environments of the atoms and bonds. At the end of the convolution steps, a pooling layer is used to generate an overall feature vector $\boldsymbol{v}_c$ for the crystal structure by taking the normalized sum of the final node embeddings $\boldsymbol{v}_i^f$. Mathematically, $\boldsymbol{v}_c$ can be written as:

\begin{align}
\boldsymbol{v}_c=\frac{1}{N} \sum_i \boldsymbol{v}_i^f, 
\end{align}

where $N$ represents the number of atoms in the unit cell of the crystal structure. This feature vector is then used as a direct input for a neural network hidden layer to predict the material property of interest which, in our study, is the formation energy.

Although iCGCNN has been shown to achieve state-of-the-art accuracy in predicting the formation energies of inorganic materials \cite{park2020developing}, such performance requires the crystal structures to be fully relaxed in terms of volume, stress, and ionic positions prior to constructing the crystal graphs. When the relaxed crystal structures are unavailable, the model performance can vary depending on how closely the unrelaxed structures resemble their respective relaxed states. To illustrate, we measured the performance of iCGCNN under four different conditions. In all conditions, the ML model was trained on 200,000 formation energy entries randomly chosen from the OQMD where the training crystal graphs were generated based on the DFT-relaxed crystal structures. However, when testing the model on another 230,000 entries, the crystal structures that were used to construct the testing crystal graphs in each condition differed in terms of their state of relaxation. In Condition \#1, fully relaxed crystal structures were used to construct the crystal graphs of the test set, while in Condition \#2, the unrelaxed crystal structures that serve as input for the DFT calculations were used. For Condition \#3, we used the crystal structures that have been relaxed in terms of volume but not in terms of stress or ionic positions. These structures were obtained by rescaling the unrelaxed crystal structures from Condition \#2 such that their volume is equivalent to that of the fully relaxed crystal structures from Condition \#1. For an additional benchmark, we trained a Magpie model \cite{ward2016general} that incorporates the Voronoi tessellation attributes \cite{ward2017including} to predict the volume of the compounds in the test set. For training, the relaxed structures were used to generate the Voronoi tessellation attributes. Voronoi tessellation attributes of the unrelaxed crystal structures were used for the volume prediction of the test data. The error of the Magpie model for predicting the volume of the crystal structures in the test data was 0.527 $\AA^3$ per atom. In Condition \#4, we rescaled the unrelaxed crystal structures from Condition \#2  to the volume predicted by the Magpie model prior to constructing the crystal graphs.

\begin{table*}
\begin{center}
\begin{tabular}{ | m{3cm} | m{3cm}| m{3cm} | m{3cm} | m{3cm} |} 
\hline
 \textbf{ML-models} &	\textbf{Condition \#1:} & \textbf{Condition \#2:} &	\textbf{Condition \#3:} &	\textbf{Condition \#4:}  \\
 &	\textbf{Fully relaxed} & \textbf{Unrelaxed} &	\textbf{Unrelaxed, but rescaled to have relaxed volume} &	\textbf{Unrelaxed, but rescaled to have volume predicted by Magpie}\cite{ward2017including}  \\
 &	\textbf{(meV/atom)} & \textbf{(meV/atom)} &	\textbf{(meV/atom)} &	\textbf{(meV/atom)}  \\
& & & & \\
\hline
iCGCNN\cite{park2020developing} & 30.1 & 62.3 & 40.2 & 49.2  \\
& & & & \\
 \hline
This work & 42.7 & 46.5 &  46.5 & 46.5 \\
 & & & & \\

\hline
\end{tabular}
\caption{\textbf{Performance test of the machine-learning models under four different conditions.} The performances of iCGCNN and the newly implemented ML model in predicting the formation energy of 230,000 materials are shown with respect to how relaxed the input crystal structures are prior to constructing the crystal graphs.}
\end{center}
\end{table*}

\begin{figure*}
\centering
\includegraphics[width=0.85\textwidth]{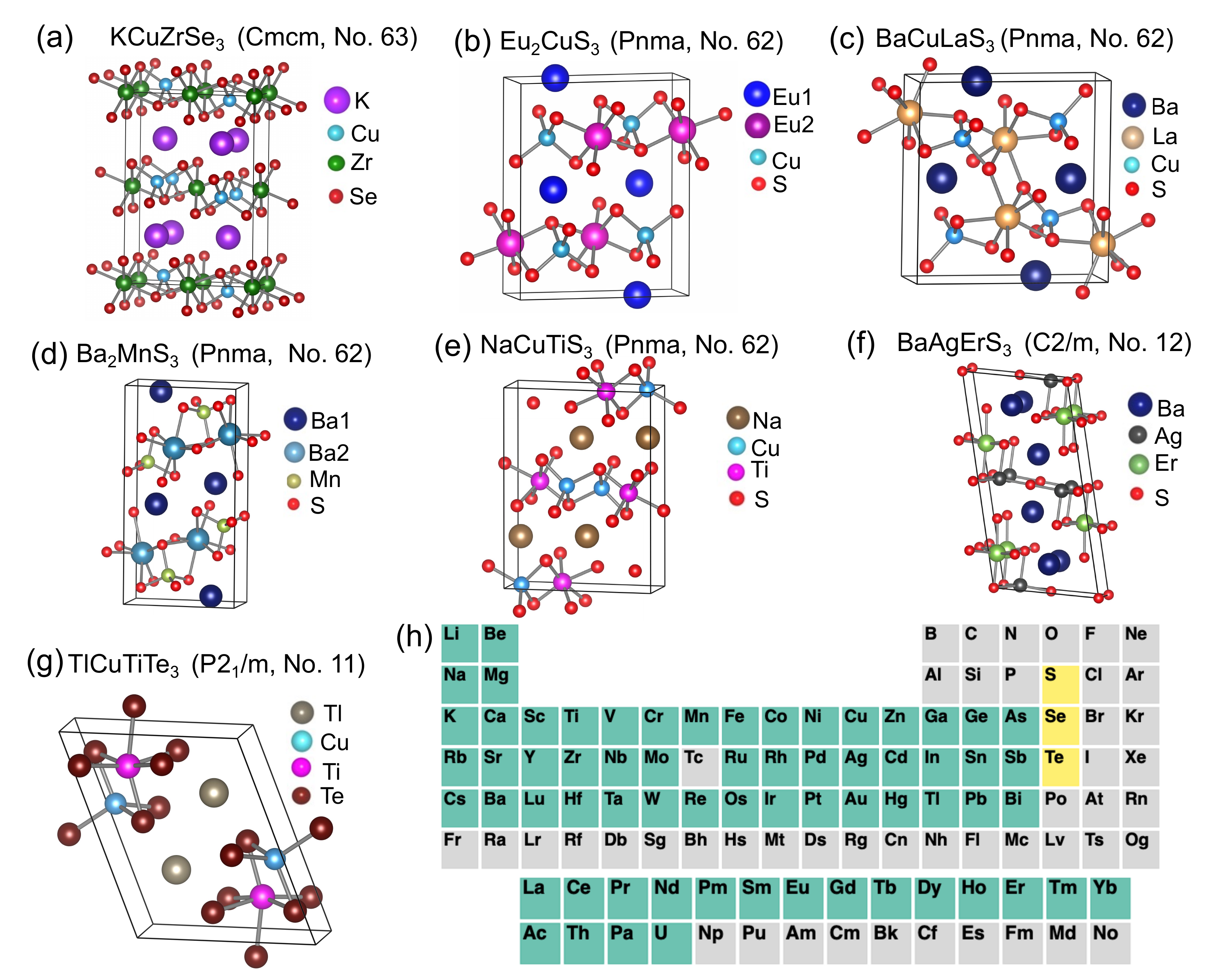}
\caption{\textbf{Structural prototypes and elements used for novel materials design.} (a-g) Conventional unit cells of the seven crystallographic prototypes that are experimentally known in the family of AMM'Q$_3$ chalcogenides are shown. We have used these crystal structures for the design and discovery of new materials. (h) We have color-coded the 66 metallic elements (highlighted in green) in periodic table that are substituted at the A, M and M' cation sites in all possible combinations to generate the initial crystal structures of the new compounds. During prototype decoration, only three chalcogen atoms (highlighted in yellow) are substituted at the Q sites.}
\end{figure*}

The performance of iCGCNN under these testing conditions are summarized in Table 1 and the relevant figures are shown in Supplementary Figure 1. Under Condition \#2, the mean absolute error (MAE) was 62.3 meV/atom. This is still significantly lower than the DFT error for experimentally measured formation energies ($\sim$ 100 meV/atom) \cite{kirklin2015open} of inorganic compounds, showing that iCGCNN remains a reliable method to predict DFT-calculated formation energies even when provided with the unrelaxed crystal structures. However, compared to the MAE of 30.1 meV/atom when the fully relaxed crystal structures were used, the error for Condition \#2 increased more than twofold indicating that the ML model is unable to perform at its peak capability when the relaxed structures are unavailable such as in a high-throughput DFT search for new materials. Under Condition \#3 and \#4, the MAE's were respectively 40.2 and 49.0 meV/atom which are 35\% and 21\% lower than that of Condition \#2. This shows that when provided with unrelaxed crystal structures as input, iCGCNN performs better the closer the volume of the structures are to their relaxed values. This further implies that iCGCNN significantly depends on the crystal volume information when predicting the formation energy of materials. 

 Here, we describe a variant iCGCNN model that independently generates the relaxed crystal volume information needed to more accurately predict the formation energy of materials when provided with the unrelaxed crystal structures as input. This is achieved through a multi-objective framework in which the ML model simultaneously predicts the relaxed volume and formation energies of the crystals. This enables the information that is used in predicting the relaxed volumes to also be utilized in predicting the formation energies and vice versa. In this improved ML  framework, before constructing the crystal graph, we normalized the crystal structures that have been provided as input for the ML model to take into account the fact that an unrelaxed crystal structure can have an arbitrary volume. The normalization process involves dividing the lattice parameters $a$, $b$, and $c$ of the unit cell by the minimum interatomic distance measured within the provided input crystal structure such that the minimum interatomic distance measured within the resulting normalized structure becomes 1. Note that during the normalization process, the fractional coordinates of the atoms with respect to the lattice vectors remain unchanged. As in iCGCNN, the normalized structure is then represented as a crystal graph where each node $n_i$ is connected to the nodes that represent the Voronoi neighbors of atom $i$. Also, node $n_i$ is embedded with vector $\boldsymbol{v}_i$ to represent the atomic properties of atom $i$, and edge ${\boldsymbol{e}_{(i,j)}}_k $ is embedded with vector ${\boldsymbol{u}_{(i,j)}}_k$ to represent the structural properties of the normalized Voronoi polyhedral formed by atoms $i$ and $j$. 

Additional to the nodes and edges, each crystal graph is associated with a scale factor s, a scalar quantity that represents the minimum interatomic distance of the crystal structure. Since the structures are normalized to have a minimum interatomic distance of 1 prior to constructing the crystal graphs, the initial value of the scale factor, $s^0$, is 1 for all crystal graphs. During the convolution steps, $s$ is iteratively updated as a function of the node embeddings. At the $t$-th convolution step, the update function for s can mathematically be written out as:

\begin{align}
f_{s}^t: s_i^{(t+1)}= \frac{1}{N} \sum_i ({{\boldsymbol{v}^{(t)}_{i}}} \boldsymbol{W}{_3}^{(t)}+\boldsymbol{b}{_3}^{(t)} )
\end{align}

where $\boldsymbol{W}_3^{(t)}$ and $\boldsymbol{b}_3^{(t)}$ represent the weight and bias of a neural network hidden layer that has a scalar output. $s$ is updated such that at the end of the convolution steps, the final value $s^f$ matches the minimum interatomic distance that would be measured in the crystal structure that has been relaxed in terms of its volume. The relaxed volume of the compound can then be predicted by simply multiplying the cube of $s^f$
to the volume of the normalized crystal structure.  

The convolution functions for updating the node and edge embeddings, $\boldsymbol{v}_i$ and ${\boldsymbol{u}_{(i,j)}}_k $ remain the same as in iCGCNN with the exception of the many-body correlation terms, ${\boldsymbol{z}^{(t)}_{(i,j)}}_k$ and ${\boldsymbol{z'}^{(t)}_{(i,j,l)}}_{(k,k')}$. For our framework, these terms are defined as ${\boldsymbol{z}^{(t)}_{(i,j)}}_k=\boldsymbol{v}_i^{(t)} \oplus \boldsymbol{v}_j^{(t)} \oplus (s^{(t)} \otimes {\boldsymbol{u}^{(t)}_{(i,j)}}_k$)  and  ${\boldsymbol{z'^{(t)}_{(i,j,l)}}}_{(k,k')}  =\boldsymbol{v}_i^{(t)} \oplus   \boldsymbol{v}_j^{(t)}  \oplus \boldsymbol{v}_l^{(t)}  \oplus (s^{(t)} \otimes {\boldsymbol{u}^{(t)}_{(i,j)}}_k)  \oplus (s^{(t)}  \otimes  {\boldsymbol{u}^{(t)}_{(i,l)}}_{k'}$)   where $\otimes$ represents a rescaling operation. In the rescaling operation, each vector element of ${\boldsymbol{u}_{(i,j)}}_k$ is multiplied by ${s^{(t)}}^d$, where the exponent $d$ represents the dimension of structural feature that is encoded in the vector element. For example, if a vector element encodes information of the area of the Voronoi surface shared by two neighboring atoms, we multiply it by  ${s^{(t)}}^2$. However, vector elements that encode the solid angle information remain unchanged during this operation. Such rescaling operation enables information used in predicting the relaxed volumes, specifically $s^{(t)}$, to be utilized by the node and edge embeddings in predicting the formation energies of materials. A schematic design of our ML model based on iCGCNN is shown in Figure 1.

\begin{figure*}
\centering
\includegraphics[width=0.9\textwidth]{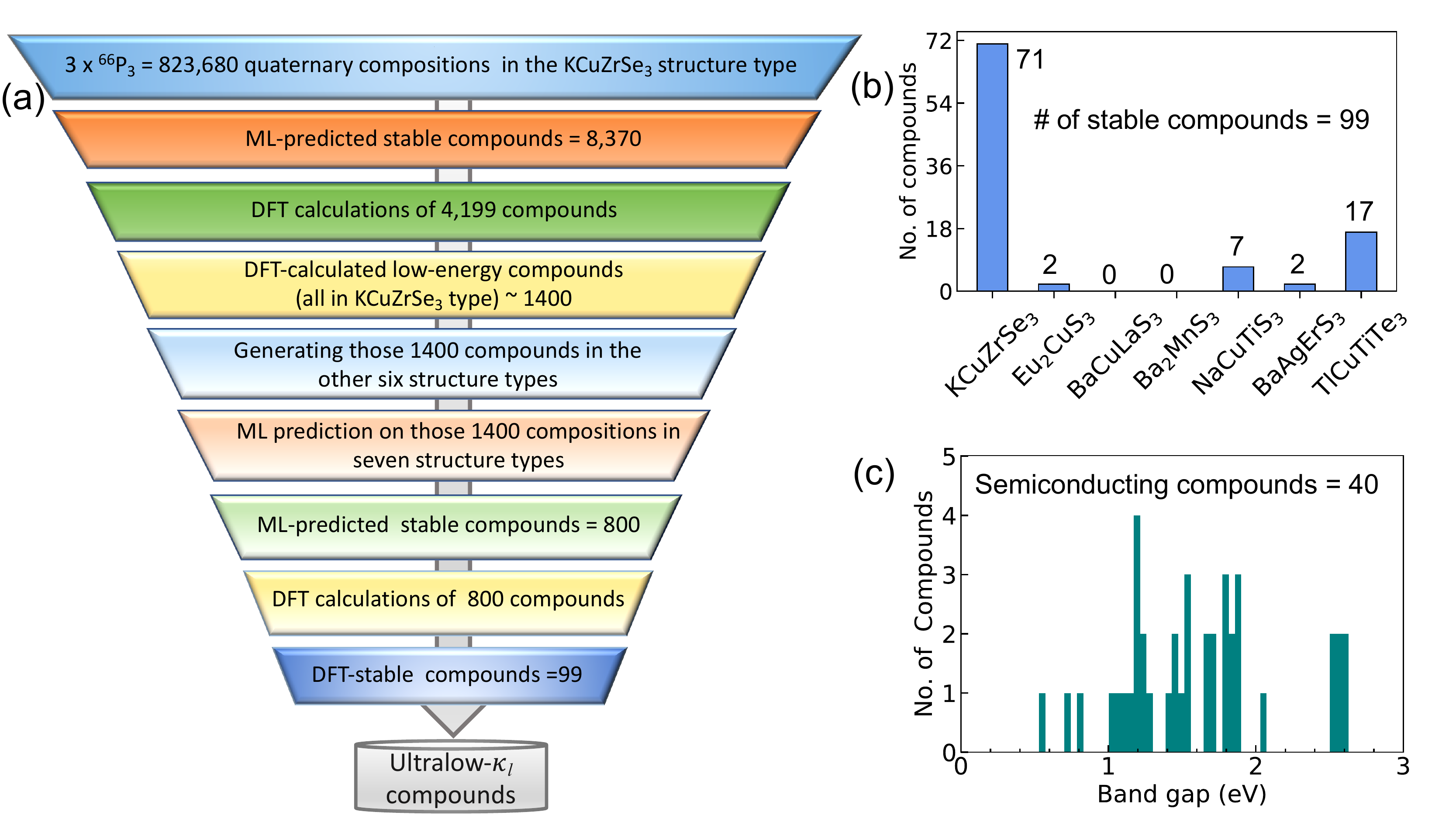}
\caption{\textbf{Machine-learning assisted discovery of novel materials} (a) Schematic work flow of novel materials discovery with iterative use of the machine learning (ML) method and density functional theory (DFT) calculations. The stable and metastable compounds are defined by their hull distance of (E$_{hd}$) = 0 and 0 $<$ E$_{hd}$ $\leq$ 50 meV/atom, respectively. (b) The distribution of predicted stable compounds into seven structure types. (c) Histogram plot showing the distribution of the band gaps of the predicted stable quaternary chalcogenides.}
\end{figure*}

Using the previous illustration, we characterize and compare the performance of our ML model with respect to iCGCNN. As shown in Table I, the MAE of our model under Condition \#1 was 42.7 meV/atom indicating that when the relaxed crystal structures are provided as input, iCGCNN outperforms the newly implemented model by 30\%. This is because our model, unlike iCGCNN, is insensitive to the volume of the input crystal structure and thus, when provided with the relaxed volume information, is unable to take advantage of it as much as iCGCNN. The same reasoning can be used to explain why iCGCNN outperforms our model by 14\% under Condition \#3. However, under Condition \#2, our model achieved an MAE of 46.5 meV/atom which is 25\% lower than iCGCNN. This shows that the new model performs significantly better than iCGCNN when the unrelaxed crystal structures are provided as input, suggesting that our model can more effectively assist in a high-throughput search for new materials even when there are fewer DFT-optimized crystal structures  available for training the model. The error of our model is also lower than that of the iCGCNN model in Condition \#4, indicating that our model improves upon existing ML methods in predicting the formation energy of materials. This improvement is most likely because the error of our model in predicting the volume of the crystal structures is 0.393 $\AA^3$ per atom, which is lower than the error of the Magpie model (0.527 $\AA^3$). The MAE of our model under Condition \#2, \#3, and \#4 are equal as expected since, by construction, the model predictions are invariant to the volume of the input crystal structures. While our model accounts for the volume differences between the relaxed and unrelaxed structures of crystal compounds, it does not account for the stress and ionic position differences that occur during relaxation. Thus, the model is expected to perform less efficiently when there is a significant difference in the unit cell shape or the ionic positions between the relaxed and unrelaxed crystal structures.

\begin{figure*}
\centering
\includegraphics[width=1.0\textwidth, trim={0cm  2cm 0cm 0cm}, clip]{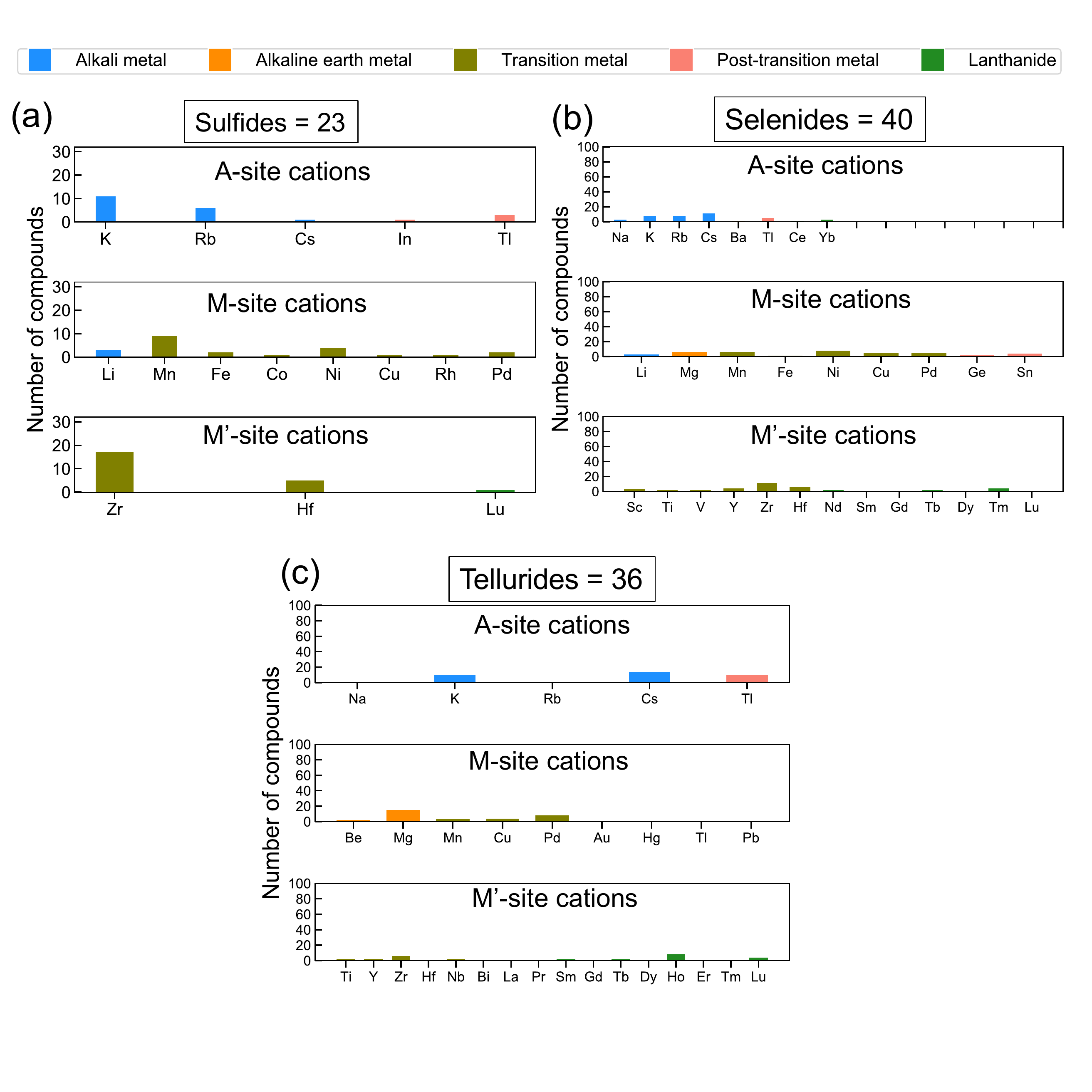}
\caption{\textbf{Elemental distributions of the predicted stable compounds.} The cations occupying the A, M and M' sites in the predicted stable quaternary chalcogenides are shown as bar charts. The height of each bar represents the number of stable compounds that contain the element. There are a total number of 99 predicted stable compounds which consists of 23 sulfides, 40 selenides and 36 tellurides.}
\end{figure*}

\begin{figure*}
\centering
\includegraphics[scale=0.6]{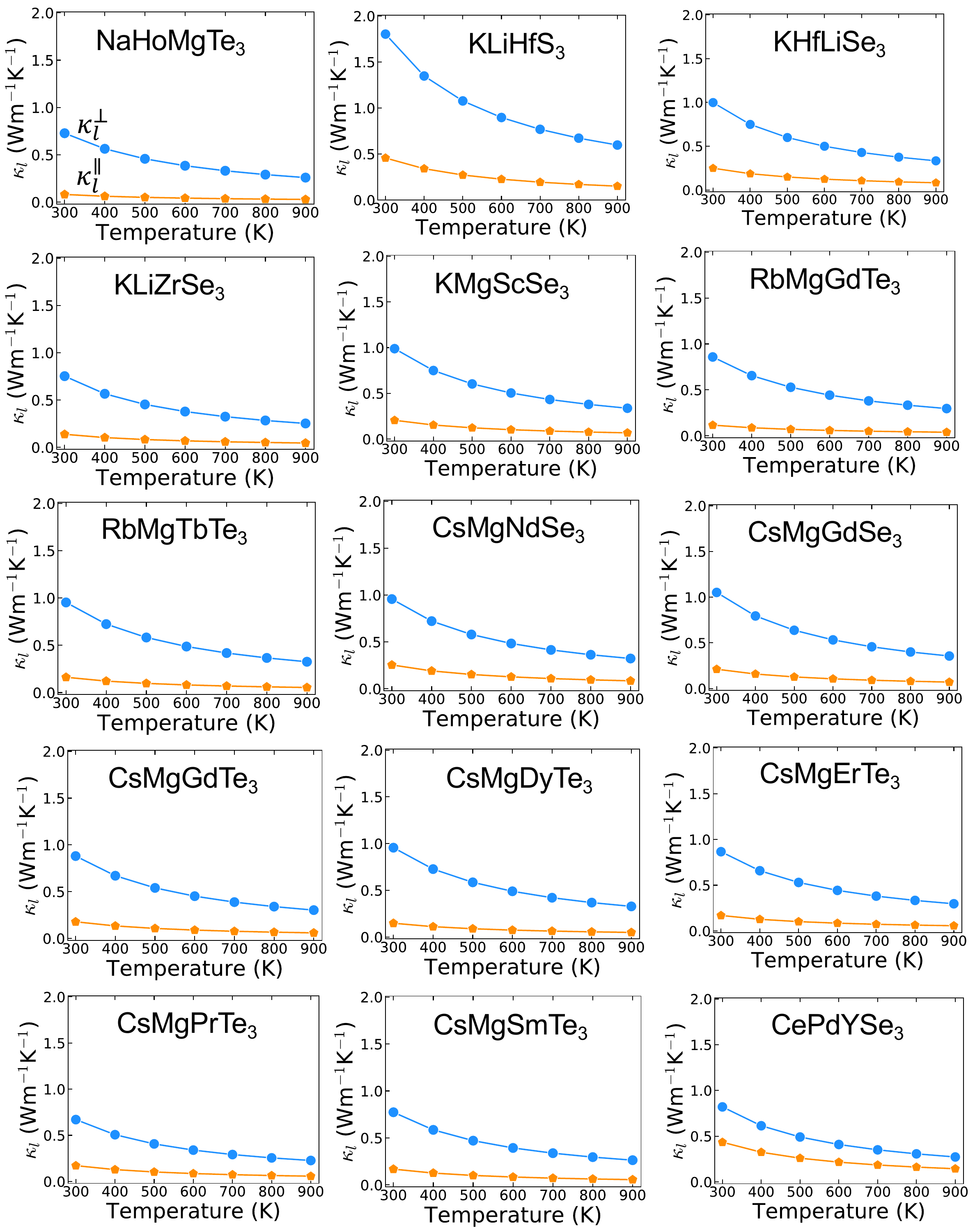}
\caption{\textbf{Thermal transport properties of the predicted stable compounds.} Calculated temperature-dependent lattice thermal conductivity ($\kappa_l$) of the 15 compounds that are selected randomly from the predicted stable and semiconducting quaternary chalcogenides. $\kappa_l^{\Vert}$ (orange pentagons) and $\kappa_l^{\bot}$ (blue circles) indicate the components of $\kappa_l$ parallel and perpendicular to the stacking direction of the layers in the crystal structure of these compounds, respectively.}
\end{figure*}

\subsection{Design and discovery of new materials}

The experimentally known AMM'Q$_3$ compounds  \cite{koscielski2012structural,strobel2006three,ruseikina2019synthesis,maier2016crystal,ruseikina2017trends,ruseikina2018refined,sikerina2007crystal,prakash2015syntheses} crystallize in seven structural prototypes (Figures 2a-2g) in four different  space groups (SGs): KCuZrSe$_3$ (SG: Cmcm, No. 63), Eu$_2$CuS$_3$ (SG: Pnma, No. 62), BaCuLaS$_3$ (SG: Pnma, No. 62), Ba$_2$MnS$_3$ (SG: Pnma, No. 62), NaCuTiS$_3$ (SG: Pnma, No. 62), BaAgErS$_3$ (SG: C2/m, No. 12), and TlCuTiTe$_3$ (SG: P2$_1$/m, No. 11) most of which are layered and related to each other via structural distortions \cite{koscielski2012structural}. Among these, KCuZrSe$_3$ has the highest symmetry that constitutes 71\% of the known AMM'Q$_3$ compounds \cite{pal2021accelerated}. We note that  while KCuZrSe$_3$, BaAgErS$_3$, and TlCuTiTe$_3$ have 12 atoms in their primitive unit cells, the rest of the structure types have 24 atoms.  In Eu$_2$CuS$_3$ and Ba$_2$MnS$_3$, Ba and Eu atoms occupy two distinct crystallographic sites in their crystal structures, respectively. We  note that the three cations (A, M, and M') occupy different crystallographic sites in the crystal structures.

First, we generate the target search space of the initial quaternary compositions taking the KCuZrSe$_3$ structure type. We chose this structure type as it is the most common in this family and all experimentally known AMM'Q$_3$ have low energies (within 50 meV/atom above the convex hull) in this prototype \cite{pal2021accelerated}. We substitute 66 metallic elements (see Figure 2h) available in the OQMD at  the K, Cu, and Zr sites in all possible combinations while keeping the Q  site  fixed only to the chalcogens (S, Se, Te). Thus, our search space  contains a total number of  $^{66}$P$_3$ $\times$ 3 = 823,680 distinct  compounds.  Next, we use the newly designed  ML model to predict 8370 stable quaternary chalcogenides with the ML-predicted hull  distances (E$_{hd}$) being  equal to zero.  In the next step,  we filter out compounds having radioactive elements and compounds which were already discovered before \cite{pal2021accelerated}. Our final set contains  4199 unique compounds in the KCuZrSe$_3$ structure type  for which we performed DFT calculations to validate the ML-predictions. After performing  T = 0K thermodynamic phase stability analysis on these compounds utilizing the data available in the OQMD, we retain $\sim$ 1400 low-energy compounds whose E$_{hd}$ lie within 50 meV/atom above the convex hull. 

In the next step, we take these 1400 compositions and generate their crystal structures in the other six structural prototypes known in the AMM'Q$_3$ family of compounds. We use the ML model again to predict stable compounds in these 1400 compositions among all structure types, which gave 800 ML-predicted stable quaternary compounds.  We perform DFT calculations on these 800 compounds and evaluate their T= 0K thermodynamic phase stability in the OQMD. Finally, we  discover  99 stable (E$_{hd}$ = 0)  and 362 low-energy metastable (0 $<$  E$_{hd}$ $\leq$ 50 meV/atom) compounds having different structure types. A schematic funnel diagram of the materials design and discovery is shown in Figure 3a. We  list these stable and  metastable compounds with their  energetic information in  the Supplementary Note.  We also note that these compounds are distinct from the compounds that were predicted in the previous HT-DFT work \cite{pal2021accelerated}  Our analysis reveals that among the 99 newly predicted stable compounds, 71 crystallize in the KCuZrSe$_3$ structure type followed by 17 compounds crystallizing in the TlCuTiTe$_3$ prototype (Figure 3b).  The other 11 compounds are found in Eu$_2$CuS$_3$ (2), NaCuTiS$_3$ (7) and BaAgErS$_3$ (2) structure types. We found no stable compounds in the BaCuLaS$_3$ and Ba$_2$MnS$_3$ prototypes. Analysis of the DFT-calculated band gap reveals that 40 of the 99 predicted stable compounds have finite band gaps that vary between 0.53 eV to 2.63 eV, which is shown as a histogram plot in Figure 3c.

We further analyze the newly predicted stable compounds  in terms of the constituent elements (see bar charts in Figure 4). The newly predicted 99 stable compounds consist of 23 sulfides, 40 selenides, and 36 tellurides. We note that the elements in all the predicted compounds are arranged to be in the A-M-M'-Q order as in  the experimentally known AMM'Q$_3$ compounds to follow their site occupancy. This helps us to identify which elements and chemical groups occupy the A, M, and M' sites.  We notice that some of the predicted stable compounds do not appear to be charge-balanced assuming the nominal oxidation states of the constituent elements e.g., NaMnZrS$_3$, NaNiTiSe$_3$, etc. We also see that the M site in some of the sulfides and selenides is occupied by alkali metal (Li) e.g., in KLiZrS$_3$ and KLiZrSe$_3$. The M-site in some of the selenides is also occupied by  alkaline earth (e.g., CsMgYSe$_3$) or post-transition metals (e.g., CsSnYSe$_3$).  Similarly, for tellurides,  the M-site is sometimes occupied by an alkaline earth metal e.g., CsMgGdTe$_3$, or  post-transition metal e.g., KPbHoTe$_3$.  The M' site in some of the tellurides is occupied by post-transition metals e.g., CsHgBiTe$_3$. It is also evident that some of the predicted stable compounds have a combination of alkali and alkaline earth metals (e.g., KMgHoTe$_3$) and more than one alkali metals (e.g., KLiHfS$_3$).  These chemical trends are unique to this newly discovered compounds and are absent in the experimentally known \cite{koscielski2012structural,strobel2006three,ruseikina2019synthesis,maier2016crystal,ruseikina2017trends,ruseikina2018refined,sikerina2007crystal,prakash2015syntheses} and the previously \cite{pal2021accelerated}  discovered AMM'Q$_3$ compounds.

%\subsection{validation using DFT}

\begin{figure*}
\centering
\includegraphics[scale=0.6, trim={3cm  0cm 1cm 0cm}, clip]{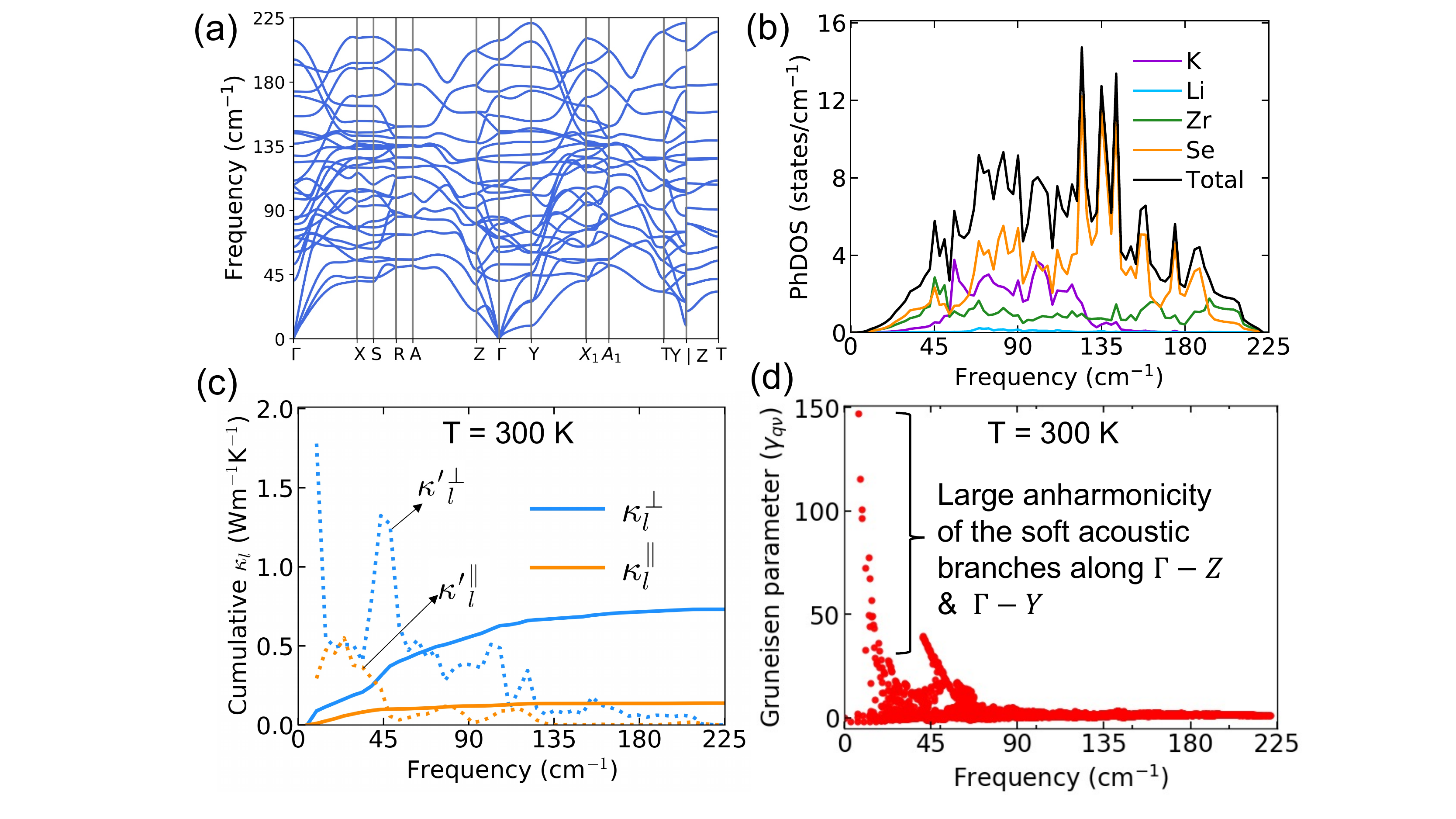}
\caption{\textbf{Harmonic and anharmonic lattice dynamics properties of KLiZrSe$_3$.} (a) Harmonic phonon dispersion and (b) density of states of KLiZrSe$_3$ shown up to 225 cm$^{-1}$.  (c) Cumulative lattice thermal conductivity plots $\kappa_l^{\bot}$ and $\kappa_l^{\Vert}$ (values are on the y-axis) and their first-order derivatives (${{\kappa_l}^{\prime}}^{\bot}$ and ${{\kappa}^{\prime}_l}^{\Vert}$, in arbitrary units) with respect to the phonon-frequency, which are obtained from the anharmonic lattice dynamics calculations. (d) Mode-Gruneisen parameters ($\gamma_{q\nu}$'s) of the  phonon frequencies (up to 225 cm$^{-1}$) obtained using the third-order IFCs.}
\end{figure*}

\subsection{Thermal transport properties}

We now  investigate the thermal transport properties of the newly discovered quaternary chalcogenides.  To this end, we focus on the  compounds which are non-magnetic and semiconducting   to  unravel how the crystal structures influence their phonon dispersions and  $\kappa_l$.  We did not choose any magnetic as well  as metallic  compounds for this purpose  as they can have significant magnonic and electronic contributions,  respectively,  to the total  thermal conductivity ($\kappa$) as opposed to semiconductors where  $\kappa_l$ dominates $\kappa$. Further, we select only those compounds that have the KCuZrSe$_3$ structure type as it has the highest symmetry and the smallest unit cell among the seven  structural prototypes known in this family of compounds. This choice makes the calculation of $\kappa_l$ computationally less expensive compared to the other structure types.   Since the calculation of $\kappa_l$ within a first-principles  framework  is computationally very  expensive, we randomly chose  a set of 15  compounds  from the list  of predicted  DFT-stable  semiconducting chalcognides and calculated their $\kappa_l$  using the phonon Boltzmann transport equation (PBTE) as detailed in the  method section. We calculate their electronic structures (see Supplementary Figure 2), phonon dispersions (see Supplementary Figure 3), and $\kappa_l$ (Figure 5).  The DFT-calculated  band gaps (E$_g$) of these selected compounds vary  from small (0.36 eV in CePdYSe$_3$)  to large (2.43 eV in CsMgGdSe$_3$)  values.  From Figure 5, it is seen that these compounds  exhibit very low $\kappa_l$ ($\kappa_l^{\bot}$ $\leq$ 1.80 Wm$^{-1}$K$^{-1}$  and $\kappa_l^{\Vert}$ $\leq$ 0.50 Wm$^{-1}$K$^{-1}$ at T $\geq$ 300 K for any compounds), which is smaller than the values reported experimentally in single-crystalline SnSe ($\kappa_l^{\bot}$ $\sim$ 1.90 Wm$^{-1}$K$^{-1}$  and $\kappa_l^{\Vert}$ $\sim$ 0.90 Wm$^{-1}$K$^{-1}$ at T = 300 K) \cite{wu2017direct}.  Here,  $\kappa_l^{\bot}$ and $\kappa_l^{\Vert}$ are the components that are parallel and perpendicular to the stacking direction in the crystal structure of the compounds, respectively. Due to the layered crystal structure, the two components of $\kappa_l$ are strongly  anisotropic with the $\kappa_l^{\bot}$  being much larger than the $\kappa_l^{\Vert}$ due to the stronger intralayer interactions.  Among these 15 compounds,    KLiHfS$_3$ has the highest ($\kappa_l^{\Vert}$: 1.80 Wm$^{-1}$K$^{-1}$, $\kappa_l^{\bot}$: 0.50 Wm$^{-1}$K$^{-1}$) and CsMgPrTe$_3$ has the lowest ($\kappa_l^{\Vert}$: 0.67 Wm$^{-1}$K$^{-1}$,  $\kappa_l^{\bot}$:  0.17 Wm$^{-1}$K$^{-1}$) value  of   $\kappa_l$  at 300 K. To understand   the origin of ultralow $ \kappa_l$ in this family of materials and to unravel the structure-property relationship,  we picked one compound KLiZrSe$_3$ to investigate its harmonic and anharmonic lattice dynamical  properties in detail.

We start the analysis of KLiZrSe$_3$ by examining its phonon dispersion and density of states which are shown up to 225 cm$^{-1}$ here. For a full phonon dispersion, see Supplementary Figure 3.  We see that the phonon dispersion (Figure 6a) exhibits (a) very soft($<$ 25 cm$^{-1}$) acoustic phonon branches along $\Gamma$-Y and $\Gamma$-Z directions, (b) several low-energy optical phonon modes near 35 cm$^{-1}$, and (b) a strong hybridization between the phonon branches up to  225 cm$^{-1}$ (Figure 6b). The calculated elastic moduli of KLiZrSe$_3$ are very low (bulk modulus: 28 GPa, Shear modulus: 20 GPa) that give rise to an average sound velocity of 2.48 km/s. While the soft elasticity and the acoustic modes lead to the low speed of sounds, the strongly hybridized low-energy phonon branches enhance the number of phonon-scattering channels, which altogether lead to an ultralow $\kappa_l$ in KLiZrSe$_3$. Figure 6a also exhibits a nearly dispersion-less optical phonon branch near 48 cm$^{-1}$ along the X-S-R-A-Z direction, which is reminiscent of rattling vibrations that also help in reducing the phonon lifetimes in this compound.  To understand the mode-wise contribution to $\kappa_l$, we plot the cumulative  values of $\kappa_l^{\Vert}$, $\kappa_l^{\bot}$, and their first-order derivatives (i.e., ${{\kappa_l}^{\prime}}^{\bot}$ and ${{\kappa}^{\prime}_l}^{\Vert}$)  as a function of the phonon frequency in KLiZrSe$_3$.  Figure 6c shows that   $\kappa_l^{\Vert}$ varies up to  45 cm$^{-1}$ and then plateaus above that frequency, indicating that only the acoustic and low-energy optical phonons up to  frequency of 45 cm$^{-1}$ mainly contribute to it. On the other hand, the variation of $\kappa_l^{\bot}$ occurs up to 135 cm$^{-1}$ before saturating to a nearly constant value. Hence, the high-energy optical phonons also contribute to $\kappa_l^{\bot}$. This anisotropy between $\kappa_l^{\Vert}$ and $\kappa_l^{\bot}$ originates from the disparate strength of intralayer and interlayer interactions in the layered crystal structure of KLiZrSe$_3$, the former being much stronger than the later.

To further enhance our understanding about  the microscopic  origin of ultralow $\kappa_l$ in KLiZrSe$_3$,  we  calculate the mode  Gruneisen parameters ($\gamma_{q\nu}$ = $\frac{dln\omega_{q\nu}}{dV}$) of the phonon modes, where $\omega_{q\nu}$ is the frequency of the phonon mode $\nu$ at the q-point and $V$ is the volume of the unit cell.  $\gamma_{q\nu}$'s   directly measure the anharmonicity of the phonon modes which play a crucial role in inducing  low-$\kappa_l$ in crystalline solids \cite{morelli2008intrinsically}.    The calculated $\gamma_{q\nu}$'s  of KLiZrSe$_3$ are very large, with the values reaching as     high as    150 for  the soft acoustic phonon modes (Figure 6d). The optical phonon frequencies (up to 80 cm$^{-1}$) also exhibit very large $\gamma_{q\nu}$'s ($>>$1). Due to such high  values of  $\gamma_{q\nu}$, the scattering  phase space of KLiZrSe$_3$ becomes very large  which leads to  an enhanced phonon-scattering events, and hence an ultralow value of $\kappa_l$ in KLiZrSe$_3$ and  this family of compounds in general. We note that $\kappa_l$ has been calculated here using only the three-phonon scattering rates. Inclusion of additional scatterings due to higher-order phonon interactions \cite{xia2020high} or grain-boundaries \cite{pal2021microscopic} may further decrease the calculated $\kappa_l$. It is interesting to note that the electronic structures (Supplementary Figure 2) of these compounds exhibit flat-and-dispersive bands as well multiple band extrema near the valence and conduction band edges, making them attractive for thermoelectric studies. 

\section{Discussion}

We construct an advanced  ML model based on the recently developed  iCGCNN model to search for novel compounds by exploring the vast composition space engulfed  by a class  of  known quaternary chalcogenides AMM'Q$_3$. The model is designed to be scale-invariant to the input crystal structure, allowing the model to predict the properties of hypothetical compounds more accurately without knowing their DFT-relaxed volumes. Using DFT as a validation tool, we  discover a  large  number  of 99 thermodynamically stable and 362 low-energy metastable compounds that are  amenable to experimental synthesis and characterization in  the laboratory. The success rate ($\sim$ 11 \%) of materials discovery is quite high in our work since we discovered a total number of 461 potentially synthesizable novel quaternary chalcogenides performing DFT calculations for 4199 unique AMM'Q$_3$ compositions.  To investigate the thermal transport properties, we randomly select  15 DFT-stable semiconducting  and non-magnetic compounds to calculate their $\kappa_l$  using the PBTE including the three-phonon scattering rates. Our calculations reveal that all of these  compounds exhibit ultralow  $\kappa_l$. By analyzing the harmonic (e.g., phonon dispersion and density of states) and anharmonic  (e.g., mode Gruneisen parameters) lattice dynamical properties of one of the  compounds KLiZrSe$_3$, we  found that  the ultralow-$\kappa_l$ in this family of compounds arises from (a) soft acoustic phonon branches that give rise to low sound velocities,  (b) the strong hybridization between  the phonons branches appearing at low-frequency, and (c) large phonon anharmonicity as evident in the very high values of the  mode-Gruniesen parameters. The two later  factors help in increasing the phonon-scattering phase  space  as well as  the phonon-scattering rates in this compound.  Furthermore, the  presence of low-energy nearly dispersion-less optical phonon branches, which  are  reminiscent of rattling phonon branches,  also  play an important role in giving rise to the small lifetime of the heat-carrying phonons, leading to a  very  low-$\kappa_l$. We  hope that our work would encourage the application and development of ML models based on graph neural network for the efficient discovery of novel materials. While our model bypasses the need to utilize the DFT-relaxed volume information of the input crystal structures, additional work must be done to design an ML model that can account for the changes that occur in crystal structures during relaxation in terms of the stress and the ionic positions. Last but not least, our results present opportunities in further experimental and  theoretical investigations of these newly discovered AMM'Q$_3$ materials having innate ultralow-$\kappa_l$ for various thermal energy management applications.

\section{Methods}
\subsection{ML code implementation}
The ML code was implemented based on the existing iCGCNN framework \cite{park2020developing}. PyTorch\cite{paszke2019pytorch} was used to implement the neural network components of our model while Pymatgen \cite{ong2013python} was used for performing the Voronoi tessellations of the crystal structures \cite{ward2017including} before constructing the crystal graphs.

\subsection{ML training data}
Our ML model was trained on DFT-calculated thermodynamic data taken from the OQMD  \cite{saal2013materials, kirklin2015open} and a previous constrained HT-DFT search  of AMM'Q$_3$-type  compounds conducted by  Pal et al. \cite{pal2021accelerated}. Approximately 430,000 unique ordered inorganic  compounds with formation energies less than 5 eV/atom were taken  from the OQMD. These include experimentally  known  compounds from the inorganic crystal structure database  (ICSD) \cite{belsky2002new} and hypothetical 
compounds  with commonly occurring structures. The HT-DFT study by Pal et al. \cite{pal2021accelerated} was  performed for 4659 compounds  in various   AMM'Q$_3$ structure types. All  4659 compounds were  included  in the ML  training data. The crystal graphs of all compounds in the training data were generated from their  relaxed  crystal structures. For all ML training in this study, 20\% of the training data were randomly chosen and reserved for validation.

\subsection{DFT calculations}
All DFT calculations were performed using the Vienna Ab-initio Simulation Package (VASP) \cite{kresse1996efficiency} employing the projector-augmented wave (PAW) \cite{blochl1994projector, kresse1999ultrasoft} potentials.  The Perdew-Burke-Ernzerhof (PBE) \cite{perdew1996generalized} generalized gradient approximation (GGA) was used to treat the   exchange and correlation energies of the electrons.  All relaxation and static calculations of the compounds for phase stability analysis  were performed in accordance with the DFT settings  as  laid out in our high-throughput  framework available with the OQMD through the \texttt{qmpy}  suit of codes \cite{saal2013materials, kirklin2015open}. 
 
\subsection{Stability analysis}
We determined the thermodynamic phase stability of the compounds by utilizing the DFT calculated total energy (T= 0 K). It has been shown that  0 K thermodynamic phase stability data can often serve as an excellent metric of the  synthesizability of a  compound   \cite{zakutayev2013theoretical, gautier2015prediction,  aykol2018thermodynamic, jain2016computational, anand2019double, hautier2012computer, curtarolo2013high, sun2017thermodynamic, kirklin2015open}. To assess the stability of a new quaternary composition we constructed its convex hull by  considering all its competing phases in that phase space. From this,  we define  the stability  of a compound by its hull distance (E$_{hd}$). For a stable  compound, by definition,  E$_{hd}$ = 0. On the other hand, a compound is considered to be metastable when its E$_{hd}$ lies within 50 meV/atom above the hull in keeping with the heuristic conventions used in literature \cite{zakutayev2013theoretical, cerqueira2015identification, wu2013first}. Compounds that have E$_{hd}$ larger than 50 meV/atom above the hull are considered to be unstable. For a detailed discussion on the convex hull construction and hull distance,  we refer to Ref. \cite{saal2013materials,kirklin2015open,jain2013commentary,curtarolo2012aflowlib, emery2016high, pal2021accelerated}. It is worth mentioning  that many known compounds in  the  ICSD are metastable with varying positive hull distances \cite{sun2016thermodynamic, aykol2018thermodynamic}.

\subsection{Calculation of lattice thermal conductivity}
We calculated  the phonon dispersions of these compounds using 2$\times$2$\times$1 supercells within the finite-displacement method as implemented in Phonopy \cite{phonopy}. The high symmetry paths in the  Brillouin zones are adopted following the conventions used by Setyawan et al. \cite{setyawan2010high} while plotting  the phonon dispersions as well the electronic structures. We calculated the lattice thermal conductivity ($\kappa_l$) utilizing the phonon life times obtained from the third order interatomic force constants (IFCs)  \cite{chaput2011phonon, togo2015distributions,zhou2014lattice}, which was  shown to reproduce  $\kappa_l$ within  5 \% of the experimentally measured $\kappa_l$  in this  AMM'Q$_3$  family of compounds \cite{hao2019design, pal2019intrinsically}. We constructed the third-order IFCs  using the compressive sensing lattice dynamics (CSLD) method \cite{zhou2014lattice,zhou2019compressive1} that utilize the displacement-force data generated from  supercell configurations. In this study we used 2$\times$2$\times$1  supercells with the cut-off radius (r$_{c}$) for the third-order IFCs taken to be the sixth nearest-neighbor distance within each crystal structure. 

Using the second and third-order  IFCs   in the ShengBTE  code \cite{li2014shengbte}, we calculated  the temperature-dependent $\kappa_l$ utilizing  an iterative solution to the phonon Boltzmann transport equation (BTE) for phonons   using a 12$\times$12$\times$12 q-point mesh.  It is known that the calculated $\kappa_l$ depends on r$_{c}$ which specifies the   maximum range of interaction in the third-order IFCs \cite{li2015orbitally} as well as on  the q-point grid.   In our earlier work \cite{pal2019intrinsically},  we showed  that  good convergence of $\kappa_l$  was obtained by  limiting  r$_c$ even to the third nearest-neighbor and with the above mesh of q-points  in this family  of   compounds.  Due to the  layered crystal structure of the AMM'Q$_3$ compounds, we present the  in-plane ($\kappa_l^{\bot}$, which is the directional  average of the $\kappa_l$ components along the two in-plane directions) and   the cross-plane ($\kappa_l^{\Vert}$, which is along the stacking direction of the crystal) components of the calculated $\kappa_l$ tensor in the result section.

\section{Data availability} The data that supports the findings of the work are in the manuscript and Supplementary Information. The structures and energetics of the predicted compounds would be made available through the Open Quantum Materials database (OQMD)  in a future release. Additional data will be available upon reasonable request.

\section{Code availability} The custom codes used in this work are available under reasonable request.

\section{Materials \& Correspondence}
Request for materials and correspondence should be addressed to K.P., C.W.P.,  or C.W.

\section{Acknowledgements} 
\begin{acknowledgements}
The authors acknowledge support from the U.S. Department of Energy under Contract No. DE-SC0014520 (thermal conductivity calculations), National  Institute of Standards and Technology as part of the  Center for Hierarchical Materials Design (CHiMaD) under the Award  70NANB14H012 by U.S. Department of Commerce (HT-DFT calculations), the Toyota  Research Institute through the Accelerated Materials Design and Discovery program (machine learning and lattice dynamics), and the National Science Foundation through the MRSEC program (NSF-DMR 1720139) at the Materials Research Center (phase stability).  We acknowledge the computing resources provided by (a) the National Energy Research Scientific Computing Center (NERSC), a U.S. Department of Energy Office of Science User Facility operated under Contract No. DE-AC02-05CH11231, (b)  Quest high-performance computing facility at Northwestern University which is jointly supported by the Office of the Provost, the Office for Research, and Northwestern University Information Technology, and (c)  the Extreme Science and Engineering Discovery Environment (National Science Foundation Contract ACI-1548562).
\end{acknowledgements}

\section{Author contributions}
K.P. conceived and designed the project. K.P. performed HT-DFT calculations and analysis with  suggestions from Y.X., J.S., and C.W. C.W.P. designed and trained the ML model, and performed ML predictions with suggestions from C.W.   C.W. supervised the whole project. All authors discussed the results, provided comments, and contributed to writing the manuscript.

\section{Competing interests} The authors declare no competing financial or non-financial interests.

%\bibliography{ml-ammq3-ref}

\begin{thebibliography}{76}%
\makeatletter
\providecommand \@ifxundefined [1]{%
 \@ifx{#1\undefined}
}%
\providecommand \@ifnum [1]{%
 \ifnum #1\expandafter \@firstoftwo
 \else \expandafter \@secondoftwo
 \fi
}%
\providecommand \@ifx [1]{%
 \ifx #1\expandafter \@firstoftwo
 \else \expandafter \@secondoftwo
 \fi
}%
\providecommand \natexlab [1]{#1}%
\providecommand \enquote  [1]{``#1''}%
\providecommand \bibnamefont  [1]{#1}%
\providecommand \bibfnamefont [1]{#1}%
\providecommand \citenamefont [1]{#1}%
\providecommand \href@noop [0]{\@secondoftwo}%
\providecommand \href [0]{\begingroup \@sanitize@url \@href}%
\providecommand \@href[1]{\@@startlink{#1}\@@href}%
\providecommand \@@href[1]{\endgroup#1\@@endlink}%
\providecommand \@sanitize@url [0]{\catcode `\\12\catcode `\$12\catcode
  `\&12\catcode `\#12\catcode `\^12\catcode `\_12\catcode `\%12\relax}%
\providecommand \@@startlink[1]{}%
\providecommand \@@endlink[0]{}%
\providecommand \url  [0]{\begingroup\@sanitize@url \@url }%
\providecommand \@url [1]{\endgroup\@href {#1}{\urlprefix }}%
\providecommand \urlprefix  [0]{URL }%
\providecommand \Eprint [0]{\href }%
\providecommand \doibase [0]{https://doi.org/}%
\providecommand \selectlanguage [0]{\@gobble}%
\providecommand \bibinfo  [0]{\@secondoftwo}%
\providecommand \bibfield  [0]{\@secondoftwo}%
\providecommand \translation [1]{[#1]}%
\providecommand \BibitemOpen [0]{}%
\providecommand \bibitemStop [0]{}%
\providecommand \bibitemNoStop [0]{.\EOS\space}%
\providecommand \EOS [0]{\spacefactor3000\relax}%
\providecommand \BibitemShut  [1]{\csname bibitem#1\endcsname}%
\let\auto@bib@innerbib\@empty
%</preamble>
\bibitem [{\citenamefont {Wu}\ \emph {et~al.}(2002)\citenamefont {Wu},
  \citenamefont {Wei}, \citenamefont {Padture}, \citenamefont {Klemens},
  \citenamefont {Gell}, \citenamefont {Garc{\'\i}a}, \citenamefont {Miranzo},\
  and\ \citenamefont {Osendi}}]{wu2002low}%
  \BibitemOpen
  \bibfield  {author} {\bibinfo {author} {\bibfnamefont {J.}~\bibnamefont
  {Wu}}, \bibinfo {author} {\bibfnamefont {X.}~\bibnamefont {Wei}}, \bibinfo
  {author} {\bibfnamefont {N.~P.}\ \bibnamefont {Padture}}, \bibinfo {author}
  {\bibfnamefont {P.~G.}\ \bibnamefont {Klemens}}, \bibinfo {author}
  {\bibfnamefont {M.}~\bibnamefont {Gell}}, \bibinfo {author} {\bibfnamefont
  {E.}~\bibnamefont {Garc{\'\i}a}}, \bibinfo {author} {\bibfnamefont
  {P.}~\bibnamefont {Miranzo}},\ and\ \bibinfo {author} {\bibfnamefont {M.~I.}\
  \bibnamefont {Osendi}},\ }\bibfield  {title} {\bibinfo {title}
  {Low-thermal-conductivity rare-earth zirconates for potential
  thermal-barrier-coating applications},\ }\href
  {https://ceramics.onlinelibrary.wiley.com/doi/abs/10.1111/j.1151-2916.2002.tb00574.x?casa_token=Ns0hPM7lvTUAAAAA:so6XjYEbwDLIR8_0-SkRfo5emfQGr1pTifkZjCKU9obFtYJTp9aExqXTYcm1w7MJg_EgSRTRoPmj1g}
  {\bibfield  {journal} {\bibinfo  {journal} {Journal of the American Ceramic
  Society}\ }\textbf {\bibinfo {volume} {85}},\ \bibinfo {pages} {3031}
  (\bibinfo {year} {2002})}\BibitemShut {NoStop}%
\bibitem [{\citenamefont {Bell}(2008)}]{bell2008cooling}%
  \BibitemOpen
  \bibfield  {author} {\bibinfo {author} {\bibfnamefont {L.~E.}\ \bibnamefont
  {Bell}},\ }\bibfield  {title} {\bibinfo {title} {Cooling, heating, generating
  power, and recovering waste heat with thermoelectric systems},\ }\href
  {https://science.sciencemag.org/content/321/5895/1457.abstract?casa_token=fmW7m_QiDcoAAAAA:U5SVcvMkb5-AAyUphc0TUkRkqdBW72Md47u7YKzosBLJW8H2AUGKuQClGCaB1GDH07FIqr32jXX-}
  {\bibfield  {journal} {\bibinfo  {journal} {Science}\ }\textbf {\bibinfo
  {volume} {321}},\ \bibinfo {pages} {1457} (\bibinfo {year}
  {2008})}\BibitemShut {NoStop}%
\bibitem [{\citenamefont {Li}\ \emph {et~al.}(2018)\citenamefont {Li},
  \citenamefont {Zheng}, \citenamefont {Lv}, \citenamefont {Liu}, \citenamefont
  {Wang}, \citenamefont {Huang}, \citenamefont {Cahill},\ and\ \citenamefont
  {Lv}}]{li2018high}%
  \BibitemOpen
  \bibfield  {author} {\bibinfo {author} {\bibfnamefont {S.}~\bibnamefont
  {Li}}, \bibinfo {author} {\bibfnamefont {Q.}~\bibnamefont {Zheng}}, \bibinfo
  {author} {\bibfnamefont {Y.}~\bibnamefont {Lv}}, \bibinfo {author}
  {\bibfnamefont {X.}~\bibnamefont {Liu}}, \bibinfo {author} {\bibfnamefont
  {X.}~\bibnamefont {Wang}}, \bibinfo {author} {\bibfnamefont {P.~Y.}\
  \bibnamefont {Huang}}, \bibinfo {author} {\bibfnamefont {D.~G.}\ \bibnamefont
  {Cahill}},\ and\ \bibinfo {author} {\bibfnamefont {B.}~\bibnamefont {Lv}},\
  }\bibfield  {title} {\bibinfo {title} {High thermal conductivity in cubic
  boron arsenide crystals},\ }\href
  {https://science.sciencemag.org/content/361/6402/579.abstract?casa_token=wLPEdiLx44wAAAAA:OEOQMWQPwbxBFQwUnNpzQe2613sIGGkFoN0nY1MNOLBdpRFxPsPFCC7AcD9TZvAzL8-mYlr0seUi}
  {\bibfield  {journal} {\bibinfo  {journal} {Science}\ }\textbf {\bibinfo
  {volume} {361}},\ \bibinfo {pages} {579} (\bibinfo {year}
  {2018})}\BibitemShut {NoStop}%
\bibitem [{\citenamefont {Lindsay}\ \emph {et~al.}(2013)\citenamefont
  {Lindsay}, \citenamefont {Broido},\ and\ \citenamefont
  {Reinecke}}]{lindsay2013first}%
  \BibitemOpen
  \bibfield  {author} {\bibinfo {author} {\bibfnamefont {L.}~\bibnamefont
  {Lindsay}}, \bibinfo {author} {\bibfnamefont {D.}~\bibnamefont {Broido}},\
  and\ \bibinfo {author} {\bibfnamefont {T.}~\bibnamefont {Reinecke}},\
  }\bibfield  {title} {\bibinfo {title} {First-principles determination of
  ultrahigh thermal conductivity of boron arsenide: A competitor for
  diamond?},\ }\href
  {https://journals.aps.org/prl/abstract/10.1103/PhysRevLett.111.025901}
  {\bibfield  {journal} {\bibinfo  {journal} {Phys. Rev. Lett.}\ }\textbf
  {\bibinfo {volume} {111}},\ \bibinfo {pages} {025901} (\bibinfo {year}
  {2013})}\BibitemShut {NoStop}%
\bibitem [{\citenamefont {Tian}\ \emph {et~al.}(2018)\citenamefont {Tian},
  \citenamefont {Song}, \citenamefont {Chen}, \citenamefont {Ravichandran},
  \citenamefont {Lv}, \citenamefont {Chen}, \citenamefont {Sullivan},
  \citenamefont {Kim}, \citenamefont {Zhou}, \citenamefont {Liu}, \citenamefont
  {Goni}, \citenamefont {Ding}, \citenamefont {Sunm}, \citenamefont {Gmage},
  \citenamefont {Sun}, \citenamefont {Ziyaee}, \citenamefont {Huyan},
  \citenamefont {Deng}, \citenamefont {Zhou}, \citenamefont {Schmidt},
  \citenamefont {Chen}, \citenamefont {Chu}, \citenamefont {Huang},
  \citenamefont {Broido}, \citenamefont {Shi}, \citenamefont {Chen},
  \citenamefont {Ren},\ and\ \citenamefont {Zhifeng}}]{tian2018unusual}%
  \BibitemOpen
  \bibfield  {author} {\bibinfo {author} {\bibfnamefont {F.}~\bibnamefont
  {Tian}}, \bibinfo {author} {\bibfnamefont {B.}~\bibnamefont {Song}}, \bibinfo
  {author} {\bibfnamefont {X.}~\bibnamefont {Chen}}, \bibinfo {author}
  {\bibfnamefont {N.~K.}\ \bibnamefont {Ravichandran}}, \bibinfo {author}
  {\bibfnamefont {Y.}~\bibnamefont {Lv}}, \bibinfo {author} {\bibfnamefont
  {K.}~\bibnamefont {Chen}}, \bibinfo {author} {\bibfnamefont {S.}~\bibnamefont
  {Sullivan}}, \bibinfo {author} {\bibfnamefont {J.}~\bibnamefont {Kim}},
  \bibinfo {author} {\bibfnamefont {Y.}~\bibnamefont {Zhou}}, \bibinfo {author}
  {\bibfnamefont {T.-H.}\ \bibnamefont {Liu}}, \bibinfo {author} {\bibfnamefont
  {M.}~\bibnamefont {Goni}}, \bibinfo {author} {\bibfnamefont {Z.}~\bibnamefont
  {Ding}}, \bibinfo {author} {\bibfnamefont {J.}~\bibnamefont {Sunm}}, \bibinfo
  {author} {\bibfnamefont {G.~A. G.~U.}\ \bibnamefont {Gmage}}, \bibinfo
  {author} {\bibfnamefont {H.}~\bibnamefont {Sun}}, \bibinfo {author}
  {\bibfnamefont {H.}~\bibnamefont {Ziyaee}}, \bibinfo {author} {\bibfnamefont
  {S.}~\bibnamefont {Huyan}}, \bibinfo {author} {\bibfnamefont
  {L.}~\bibnamefont {Deng}}, \bibinfo {author} {\bibfnamefont {J.}~\bibnamefont
  {Zhou}}, \bibinfo {author} {\bibfnamefont {A.~J.}\ \bibnamefont {Schmidt}},
  \bibinfo {author} {\bibfnamefont {S.}~\bibnamefont {Chen}}, \bibinfo {author}
  {\bibfnamefont {C.-W.}\ \bibnamefont {Chu}}, \bibinfo {author} {\bibfnamefont
  {P.}~\bibnamefont {Huang}}, \bibinfo {author} {\bibfnamefont
  {D.}~\bibnamefont {Broido}}, \bibinfo {author} {\bibfnamefont
  {L.}~\bibnamefont {Shi}}, \bibinfo {author} {\bibfnamefont {G.}~\bibnamefont
  {Chen}}, \bibinfo {author} {\bibnamefont {Ren}},\ and\ \bibinfo {author}
  {\bibnamefont {Zhifeng}},\ }\bibfield  {title} {\bibinfo {title} {Unusual
  high thermal conductivity in boron arsenide bulk crystals},\ }\href
  {https://science.sciencemag.org/content/361/6402/582.abstract?casa_token=1GTSfjcO6rQAAAAA:YctCnSxexTp9k6gvaSKBLuoQtWnXZPaJIVC12yqIrPZP_Ykae3BbRG4dLiQVMO797hAX3vQGSGIC}
  {\bibfield  {journal} {\bibinfo  {journal} {Science}\ }\textbf {\bibinfo
  {volume} {361}},\ \bibinfo {pages} {582} (\bibinfo {year}
  {2018})}\BibitemShut {NoStop}%
\bibitem [{\citenamefont {Samanta}\ \emph {et~al.}(2020)\citenamefont
  {Samanta}, \citenamefont {Pal}, \citenamefont {Waghmare},\ and\ \citenamefont
  {Biswas}}]{samanta2020intrinsically}%
  \BibitemOpen
  \bibfield  {author} {\bibinfo {author} {\bibfnamefont {M.}~\bibnamefont
  {Samanta}}, \bibinfo {author} {\bibfnamefont {K.}~\bibnamefont {Pal}},
  \bibinfo {author} {\bibfnamefont {U.~V.}\ \bibnamefont {Waghmare}},\ and\
  \bibinfo {author} {\bibfnamefont {K.}~\bibnamefont {Biswas}},\ }\bibfield
  {title} {\bibinfo {title} {Intrinsically low thermal conductivity and high
  carrier mobility in dual topological quantum material, n-type bite},\ }\href
  {https://doi.org/10.1002/ange.202000343} {\bibfield  {journal} {\bibinfo
  {journal} {Angew. Chem.}\ }\textbf {\bibinfo {volume} {132}},\ \bibinfo
  {pages} {4852} (\bibinfo {year} {2020})}\BibitemShut {NoStop}%
\bibitem [{\citenamefont {Mukhopadhyay}\ \emph {et~al.}(2018)\citenamefont
  {Mukhopadhyay}, \citenamefont {Parker}, \citenamefont {Sales}, \citenamefont
  {Puretzky}, \citenamefont {McGuire},\ and\ \citenamefont
  {Lindsay}}]{mukhopadhyay2018two}%
  \BibitemOpen
  \bibfield  {author} {\bibinfo {author} {\bibfnamefont {S.}~\bibnamefont
  {Mukhopadhyay}}, \bibinfo {author} {\bibfnamefont {D.~S.}\ \bibnamefont
  {Parker}}, \bibinfo {author} {\bibfnamefont {B.~C.}\ \bibnamefont {Sales}},
  \bibinfo {author} {\bibfnamefont {A.~A.}\ \bibnamefont {Puretzky}}, \bibinfo
  {author} {\bibfnamefont {M.~A.}\ \bibnamefont {McGuire}},\ and\ \bibinfo
  {author} {\bibfnamefont {L.}~\bibnamefont {Lindsay}},\ }\bibfield  {title}
  {\bibinfo {title} {Two-channel model for ultralow thermal conductivity of
  crystalline tl3vse4},\ }\href
  {https://science.sciencemag.org/content/360/6396/1455.abstract?casa_token=pxoU8ZkqfzQAAAAA:QC9k_GEJxj5AD-G8nI_UyL3Gv8f6cq8cfxnjprOwqSa3JWrrxSH8dlDove9L6wXWBwm8OZtw5Dg}
  {\bibfield  {journal} {\bibinfo  {journal} {Science}\ }\textbf {\bibinfo
  {volume} {360}},\ \bibinfo {pages} {1455} (\bibinfo {year}
  {2018})}\BibitemShut {NoStop}%
\bibitem [{\citenamefont {Xia}\ \emph {et~al.}(2020{\natexlab{a}})\citenamefont
  {Xia}, \citenamefont {Pal}, \citenamefont {He}, \citenamefont
  {Ozoli{\c{n}}{\v{s}}},\ and\ \citenamefont
  {Wolverton}}]{xia2020particlelike}%
  \BibitemOpen
  \bibfield  {author} {\bibinfo {author} {\bibfnamefont {Y.}~\bibnamefont
  {Xia}}, \bibinfo {author} {\bibfnamefont {K.}~\bibnamefont {Pal}}, \bibinfo
  {author} {\bibfnamefont {J.}~\bibnamefont {He}}, \bibinfo {author}
  {\bibfnamefont {V.}~\bibnamefont {Ozoli{\c{n}}{\v{s}}}},\ and\ \bibinfo
  {author} {\bibfnamefont {C.}~\bibnamefont {Wolverton}},\ }\bibfield  {title}
  {\bibinfo {title} {Particlelike phonon propagation dominates ultralow lattice
  thermal conductivity in crystalline tl3vse4},\ }\href
  {https://journals.aps.org/prl/abstract/10.1103/PhysRevLett.124.065901}
  {\bibfield  {journal} {\bibinfo  {journal} {Phys. Rev. Lett.}\ }\textbf
  {\bibinfo {volume} {124}},\ \bibinfo {pages} {065901} (\bibinfo {year}
  {2020}{\natexlab{a}})}\BibitemShut {NoStop}%
\bibitem [{\citenamefont {Biswas}\ \emph {et~al.}(2012)\citenamefont {Biswas},
  \citenamefont {He}, \citenamefont {Blum}, \citenamefont {Wu}, \citenamefont
  {Hogan}, \citenamefont {Seidman}, \citenamefont {Dravid},\ and\ \citenamefont
  {Kanatzidis}}]{biswas2012high}%
  \BibitemOpen
  \bibfield  {author} {\bibinfo {author} {\bibfnamefont {K.}~\bibnamefont
  {Biswas}}, \bibinfo {author} {\bibfnamefont {J.}~\bibnamefont {He}}, \bibinfo
  {author} {\bibfnamefont {I.~D.}\ \bibnamefont {Blum}}, \bibinfo {author}
  {\bibfnamefont {C.-I.}\ \bibnamefont {Wu}}, \bibinfo {author} {\bibfnamefont
  {T.~P.}\ \bibnamefont {Hogan}}, \bibinfo {author} {\bibfnamefont {D.~N.}\
  \bibnamefont {Seidman}}, \bibinfo {author} {\bibfnamefont {V.~P.}\
  \bibnamefont {Dravid}},\ and\ \bibinfo {author} {\bibfnamefont {M.~G.}\
  \bibnamefont {Kanatzidis}},\ }\bibfield  {title} {\bibinfo {title}
  {High-performance bulk thermoelectrics with all-scale hierarchical
  architectures},\ }\href@noop {} {\bibfield  {journal} {\bibinfo  {journal}
  {Nature}\ }\textbf {\bibinfo {volume} {489}},\ \bibinfo {pages} {414}
  (\bibinfo {year} {2012})}\BibitemShut {NoStop}%
\bibitem [{\citenamefont {Zhao}\ \emph {et~al.}(2014)\citenamefont {Zhao},
  \citenamefont {Lo}, \citenamefont {Zhang}, \citenamefont {Sun}, \citenamefont
  {Tan}, \citenamefont {Uher}, \citenamefont {Wolverton}, \citenamefont
  {Dravid},\ and\ \citenamefont {Kanatzidis}}]{zhao2014ultralow}%
  \BibitemOpen
  \bibfield  {author} {\bibinfo {author} {\bibfnamefont {L.-D.}\ \bibnamefont
  {Zhao}}, \bibinfo {author} {\bibfnamefont {S.-H.}\ \bibnamefont {Lo}},
  \bibinfo {author} {\bibfnamefont {Y.}~\bibnamefont {Zhang}}, \bibinfo
  {author} {\bibfnamefont {H.}~\bibnamefont {Sun}}, \bibinfo {author}
  {\bibfnamefont {G.}~\bibnamefont {Tan}}, \bibinfo {author} {\bibfnamefont
  {C.}~\bibnamefont {Uher}}, \bibinfo {author} {\bibfnamefont {C.}~\bibnamefont
  {Wolverton}}, \bibinfo {author} {\bibfnamefont {V.~P.}\ \bibnamefont
  {Dravid}},\ and\ \bibinfo {author} {\bibfnamefont {M.~G.}\ \bibnamefont
  {Kanatzidis}},\ }\bibfield  {title} {\bibinfo {title} {Ultralow thermal
  conductivity and high thermoelectric figure of merit in snse crystals},\
  }\href {https://www.nature.com/articles/nature13184/} {\bibfield  {journal}
  {\bibinfo  {journal} {Nature}\ }\textbf {\bibinfo {volume} {508}},\ \bibinfo
  {pages} {373} (\bibinfo {year} {2014})}\BibitemShut {NoStop}%
\bibitem [{\citenamefont {Slade}\ \emph {et~al.}(2020)\citenamefont {Slade},
  \citenamefont {Pal}, \citenamefont {Grovogui}, \citenamefont {Bailey},
  \citenamefont {Male}, \citenamefont {Khoury}, \citenamefont {Zhou},
  \citenamefont {Chung}, \citenamefont {Snyder}, \citenamefont {Uher} \emph
  {et~al.}}]{slade2020contrasting}%
  \BibitemOpen
  \bibfield  {author} {\bibinfo {author} {\bibfnamefont {T.~J.}\ \bibnamefont
  {Slade}}, \bibinfo {author} {\bibfnamefont {K.}~\bibnamefont {Pal}}, \bibinfo
  {author} {\bibfnamefont {J.~A.}\ \bibnamefont {Grovogui}}, \bibinfo {author}
  {\bibfnamefont {T.~P.}\ \bibnamefont {Bailey}}, \bibinfo {author}
  {\bibfnamefont {J.}~\bibnamefont {Male}}, \bibinfo {author} {\bibfnamefont
  {J.~F.}\ \bibnamefont {Khoury}}, \bibinfo {author} {\bibfnamefont
  {X.}~\bibnamefont {Zhou}}, \bibinfo {author} {\bibfnamefont {D.~Y.}\
  \bibnamefont {Chung}}, \bibinfo {author} {\bibfnamefont {G.~J.}\ \bibnamefont
  {Snyder}}, \bibinfo {author} {\bibfnamefont {C.}~\bibnamefont {Uher}}, \emph
  {et~al.},\ }\bibfield  {title} {\bibinfo {title} {Contrasting snte--nasbte2
  and snte--nabite2 thermoelectric alloys: High performance facilitated by
  increased cation vacancies and lattice softening},\ }\href
  {https://pubs.acs.org/doi/abs/10.1021/jacs.0c05650} {\bibfield  {journal}
  {\bibinfo  {journal} {Journal of the American Chemical Society}\ }\textbf
  {\bibinfo {volume} {142}},\ \bibinfo {pages} {12524} (\bibinfo {year}
  {2020})}\BibitemShut {NoStop}%
\bibitem [{\citenamefont {Curtarolo}\ \emph {et~al.}(2013)\citenamefont
  {Curtarolo}, \citenamefont {Hart}, \citenamefont {Nardelli}, \citenamefont
  {Mingo}, \citenamefont {Sanvito},\ and\ \citenamefont
  {Levy}}]{curtarolo2013high}%
  \BibitemOpen
  \bibfield  {author} {\bibinfo {author} {\bibfnamefont {S.}~\bibnamefont
  {Curtarolo}}, \bibinfo {author} {\bibfnamefont {G.~L.}\ \bibnamefont {Hart}},
  \bibinfo {author} {\bibfnamefont {M.~B.}\ \bibnamefont {Nardelli}}, \bibinfo
  {author} {\bibfnamefont {N.}~\bibnamefont {Mingo}}, \bibinfo {author}
  {\bibfnamefont {S.}~\bibnamefont {Sanvito}},\ and\ \bibinfo {author}
  {\bibfnamefont {O.}~\bibnamefont {Levy}},\ }\bibfield  {title} {\bibinfo
  {title} {The high-throughput highway to computational materials design},\
  }\href
  {https://idp.nature.com/authorize/casa?redirect_uri=https://www.nature.com/articles/nmat3568&casa_token=iEO1ujcKPlEAAAAA:O1zr3dT2OPED4_B0iw8N5N96Io7oK4QtAAm9LSGapQddI-0-AW2XTWqlyGWWH-o8fmcB83mikD4pjg}
  {\bibfield  {journal} {\bibinfo  {journal} {Nature materials}\ }\textbf
  {\bibinfo {volume} {12}},\ \bibinfo {pages} {191} (\bibinfo {year}
  {2013})}\BibitemShut {NoStop}%
\bibitem [{\citenamefont {Jain}\ \emph {et~al.}(2016)\citenamefont {Jain},
  \citenamefont {Shin},\ and\ \citenamefont {Persson}}]{jain2016computational}%
  \BibitemOpen
  \bibfield  {author} {\bibinfo {author} {\bibfnamefont {A.}~\bibnamefont
  {Jain}}, \bibinfo {author} {\bibfnamefont {Y.}~\bibnamefont {Shin}},\ and\
  \bibinfo {author} {\bibfnamefont {K.~A.}\ \bibnamefont {Persson}},\
  }\bibfield  {title} {\bibinfo {title} {Computational predictions of energy
  materials using density functional theory},\ }\href
  {https://idp.nature.com/authorize/casa?redirect_uri=https://www.nature.com/articles/natrevmats20154&casa_token=SGo6794YCFAAAAAA:bG-xBi_Nh0s9_kDJ2loJpmyu_PqneDNuD5ZEvakgRGZDYJfR-cWn33kH-hPhdHV4fsetm5YuRiVxfg}
  {\bibfield  {journal} {\bibinfo  {journal} {Nature Reviews Materials}\
  }\textbf {\bibinfo {volume} {1}},\ \bibinfo {pages} {1} (\bibinfo {year}
  {2016})}\BibitemShut {NoStop}%
\bibitem [{\citenamefont {Saal}\ \emph {et~al.}(2013)\citenamefont {Saal},
  \citenamefont {Kirklin}, \citenamefont {Aykol}, \citenamefont {Meredig},\
  and\ \citenamefont {Wolverton}}]{saal2013materials}%
  \BibitemOpen
  \bibfield  {author} {\bibinfo {author} {\bibfnamefont {J.~E.}\ \bibnamefont
  {Saal}}, \bibinfo {author} {\bibfnamefont {S.}~\bibnamefont {Kirklin}},
  \bibinfo {author} {\bibfnamefont {M.}~\bibnamefont {Aykol}}, \bibinfo
  {author} {\bibfnamefont {B.}~\bibnamefont {Meredig}},\ and\ \bibinfo {author}
  {\bibfnamefont {C.}~\bibnamefont {Wolverton}},\ }\bibfield  {title} {\bibinfo
  {title} {Materials design and discovery with high-throughput density
  functional theory: the open quantum materials database (oqmd)},\ }\href
  {https://link.springer.com/article/10.1007/s11837-013-0755-4} {\bibfield
  {journal} {\bibinfo  {journal} {Jom}\ }\textbf {\bibinfo {volume} {65}},\
  \bibinfo {pages} {1501} (\bibinfo {year} {2013})}\BibitemShut {NoStop}%
\bibitem [{\citenamefont {Jain}\ \emph {et~al.}(2013)\citenamefont {Jain},
  \citenamefont {Ong}, \citenamefont {Hautier}, \citenamefont {Chen},
  \citenamefont {Richards}, \citenamefont {Dacek}, \citenamefont {Cholia},
  \citenamefont {Gunter}, \citenamefont {Skinner}, \citenamefont {Ceder} \emph
  {et~al.}}]{jain2013commentary}%
  \BibitemOpen
  \bibfield  {author} {\bibinfo {author} {\bibfnamefont {A.}~\bibnamefont
  {Jain}}, \bibinfo {author} {\bibfnamefont {S.~P.}\ \bibnamefont {Ong}},
  \bibinfo {author} {\bibfnamefont {G.}~\bibnamefont {Hautier}}, \bibinfo
  {author} {\bibfnamefont {W.}~\bibnamefont {Chen}}, \bibinfo {author}
  {\bibfnamefont {W.~D.}\ \bibnamefont {Richards}}, \bibinfo {author}
  {\bibfnamefont {S.}~\bibnamefont {Dacek}}, \bibinfo {author} {\bibfnamefont
  {S.}~\bibnamefont {Cholia}}, \bibinfo {author} {\bibfnamefont
  {D.}~\bibnamefont {Gunter}}, \bibinfo {author} {\bibfnamefont
  {D.}~\bibnamefont {Skinner}}, \bibinfo {author} {\bibfnamefont
  {G.}~\bibnamefont {Ceder}}, \emph {et~al.},\ }\bibfield  {title} {\bibinfo
  {title} {Commentary: The materials project: A materials genome approach to
  accelerating materials innovation},\ }\href
  {https://aip.scitation.org/doi/abs/10.1063/1.4812323} {\bibfield  {journal}
  {\bibinfo  {journal} {Apl Materials}\ }\textbf {\bibinfo {volume} {1}},\
  \bibinfo {pages} {011002} (\bibinfo {year} {2013})}\BibitemShut {NoStop}%
\bibitem [{\citenamefont {Curtarolo}\ \emph {et~al.}(2012)\citenamefont
  {Curtarolo}, \citenamefont {Setyawan}, \citenamefont {Wang}, \citenamefont
  {Xue}, \citenamefont {Yang}, \citenamefont {Taylor}, \citenamefont {Nelson},
  \citenamefont {Hart}, \citenamefont {Sanvito}, \citenamefont
  {Buongiorno-Nardelli} \emph {et~al.}}]{curtarolo2012aflowlib}%
  \BibitemOpen
  \bibfield  {author} {\bibinfo {author} {\bibfnamefont {S.}~\bibnamefont
  {Curtarolo}}, \bibinfo {author} {\bibfnamefont {W.}~\bibnamefont {Setyawan}},
  \bibinfo {author} {\bibfnamefont {S.}~\bibnamefont {Wang}}, \bibinfo {author}
  {\bibfnamefont {J.}~\bibnamefont {Xue}}, \bibinfo {author} {\bibfnamefont
  {K.}~\bibnamefont {Yang}}, \bibinfo {author} {\bibfnamefont {R.~H.}\
  \bibnamefont {Taylor}}, \bibinfo {author} {\bibfnamefont {L.~J.}\
  \bibnamefont {Nelson}}, \bibinfo {author} {\bibfnamefont {G.~L.}\
  \bibnamefont {Hart}}, \bibinfo {author} {\bibfnamefont {S.}~\bibnamefont
  {Sanvito}}, \bibinfo {author} {\bibfnamefont {M.}~\bibnamefont
  {Buongiorno-Nardelli}}, \emph {et~al.},\ }\bibfield  {title} {\bibinfo
  {title} {Aflowlib. org: A distributed materials properties repository from
  high-throughput ab initio calculations},\ }\href
  {https://www.sciencedirect.com/science/article/pii/S0927025612000687?casa_token=76XVuO_yvxQAAAAA:MXxrMN2gtSlMWk_BPpgoQSCYT8Gr84E82F5u3ZskgaeKYXMK1RzcwzOSAaCEl2mEcfwm8-4}
  {\bibfield  {journal} {\bibinfo  {journal} {Computational Materials Science}\
  }\textbf {\bibinfo {volume} {58}},\ \bibinfo {pages} {227} (\bibinfo {year}
  {2012})}\BibitemShut {NoStop}%
\bibitem [{\citenamefont {Kirklin}\ \emph {et~al.}(2015)\citenamefont
  {Kirklin}, \citenamefont {Saal}, \citenamefont {Meredig}, \citenamefont
  {Thompson}, \citenamefont {Doak}, \citenamefont {Aykol}, \citenamefont
  {R{\"u}hl},\ and\ \citenamefont {Wolverton}}]{kirklin2015open}%
  \BibitemOpen
  \bibfield  {author} {\bibinfo {author} {\bibfnamefont {S.}~\bibnamefont
  {Kirklin}}, \bibinfo {author} {\bibfnamefont {J.~E.}\ \bibnamefont {Saal}},
  \bibinfo {author} {\bibfnamefont {B.}~\bibnamefont {Meredig}}, \bibinfo
  {author} {\bibfnamefont {A.}~\bibnamefont {Thompson}}, \bibinfo {author}
  {\bibfnamefont {J.~W.}\ \bibnamefont {Doak}}, \bibinfo {author}
  {\bibfnamefont {M.}~\bibnamefont {Aykol}}, \bibinfo {author} {\bibfnamefont
  {S.}~\bibnamefont {R{\"u}hl}},\ and\ \bibinfo {author} {\bibfnamefont
  {C.}~\bibnamefont {Wolverton}},\ }\bibfield  {title} {\bibinfo {title} {The
  open quantum materials database (oqmd): assessing the accuracy of dft
  formation energies},\ }\href
  {https://www.nature.com/articles/npjcompumats201510} {\bibfield  {journal}
  {\bibinfo  {journal} {npj Computational Materials}\ }\textbf {\bibinfo
  {volume} {1}},\ \bibinfo {pages} {1} (\bibinfo {year} {2015})}\BibitemShut
  {NoStop}%
\bibitem [{\citenamefont {Rupp}(2015)}]{rupp2015machine}%
  \BibitemOpen
  \bibfield  {author} {\bibinfo {author} {\bibfnamefont {M.}~\bibnamefont
  {Rupp}},\ }\bibfield  {title} {\bibinfo {title} {Machine learning for quantum
  mechanics in a nutshell},\ }\href
  {https://onlinelibrary.wiley.com/doi/abs/10.1002/qua.24954} {\bibfield
  {journal} {\bibinfo  {journal} {International Journal of Quantum Chemistry}\
  }\textbf {\bibinfo {volume} {115}},\ \bibinfo {pages} {1058} (\bibinfo {year}
  {2015})}\BibitemShut {NoStop}%
\bibitem [{\citenamefont {Ward}\ \emph {et~al.}(2016)\citenamefont {Ward},
  \citenamefont {Agrawal}, \citenamefont {Choudhary},\ and\ \citenamefont
  {Wolverton}}]{ward2016general}%
  \BibitemOpen
  \bibfield  {author} {\bibinfo {author} {\bibfnamefont {L.}~\bibnamefont
  {Ward}}, \bibinfo {author} {\bibfnamefont {A.}~\bibnamefont {Agrawal}},
  \bibinfo {author} {\bibfnamefont {A.}~\bibnamefont {Choudhary}},\ and\
  \bibinfo {author} {\bibfnamefont {C.}~\bibnamefont {Wolverton}},\ }\bibfield
  {title} {\bibinfo {title} {A general-purpose machine learning framework for
  predicting properties of inorganic materials},\ }\href
  {https://www.nature.com/articles/npjcompumats201628?report=reader} {\bibfield
   {journal} {\bibinfo  {journal} {npj Computational Materials}\ }\textbf
  {\bibinfo {volume} {2}},\ \bibinfo {pages} {16028} (\bibinfo {year}
  {2016})}\BibitemShut {NoStop}%
\bibitem [{\citenamefont {Faber}\ \emph {et~al.}(2016)\citenamefont {Faber},
  \citenamefont {Lindmaa}, \citenamefont {Von~Lilienfeld},\ and\ \citenamefont
  {Armiento}}]{faber2016machine}%
  \BibitemOpen
  \bibfield  {author} {\bibinfo {author} {\bibfnamefont {F.~A.}\ \bibnamefont
  {Faber}}, \bibinfo {author} {\bibfnamefont {A.}~\bibnamefont {Lindmaa}},
  \bibinfo {author} {\bibfnamefont {O.~A.}\ \bibnamefont {Von~Lilienfeld}},\
  and\ \bibinfo {author} {\bibfnamefont {R.}~\bibnamefont {Armiento}},\
  }\bibfield  {title} {\bibinfo {title} {Machine learning energies of 2 million
  elpasolite (a b c 2 d 6) crystals},\ }\href
  {https://journals.aps.org/prl/abstract/10.1103/PhysRevLett.117.135502}
  {\bibfield  {journal} {\bibinfo  {journal} {Physical Review Letters}\
  }\textbf {\bibinfo {volume} {117}},\ \bibinfo {pages} {135502} (\bibinfo
  {year} {2016})}\BibitemShut {NoStop}%
\bibitem [{\citenamefont {Faber}\ \emph {et~al.}(2015)\citenamefont {Faber},
  \citenamefont {Lindmaa}, \citenamefont {von Lilienfeld},\ and\ \citenamefont
  {Armiento}}]{faber2015crystal}%
  \BibitemOpen
  \bibfield  {author} {\bibinfo {author} {\bibfnamefont {F.}~\bibnamefont
  {Faber}}, \bibinfo {author} {\bibfnamefont {A.}~\bibnamefont {Lindmaa}},
  \bibinfo {author} {\bibfnamefont {O.~A.}\ \bibnamefont {von Lilienfeld}},\
  and\ \bibinfo {author} {\bibfnamefont {R.}~\bibnamefont {Armiento}},\
  }\bibfield  {title} {\bibinfo {title} {Crystal structure representations for
  machine learning models of formation energies},\ }\href
  {https://onlinelibrary.wiley.com/doi/abs/10.1002/qua.24917?casa_token=6c0qu5UwLIkAAAAA:JwzV-gxjtMjDg4MLWX1be15Z_OT4_dKLRteX1UTpZdDnY7d5o2bIAl5-4HeWauIsEzb4D8xvNSnYsQ}
  {\bibfield  {journal} {\bibinfo  {journal} {International Journal of Quantum
  Chemistry}\ }\textbf {\bibinfo {volume} {115}},\ \bibinfo {pages} {1094}
  (\bibinfo {year} {2015})}\BibitemShut {NoStop}%
\bibitem [{\citenamefont {Hautier}\ \emph {et~al.}(2010)\citenamefont
  {Hautier}, \citenamefont {Fischer}, \citenamefont {Jain}, \citenamefont
  {Mueller},\ and\ \citenamefont {Ceder}}]{hautier2010finding}%
  \BibitemOpen
  \bibfield  {author} {\bibinfo {author} {\bibfnamefont {G.}~\bibnamefont
  {Hautier}}, \bibinfo {author} {\bibfnamefont {C.~C.}\ \bibnamefont
  {Fischer}}, \bibinfo {author} {\bibfnamefont {A.}~\bibnamefont {Jain}},
  \bibinfo {author} {\bibfnamefont {T.}~\bibnamefont {Mueller}},\ and\ \bibinfo
  {author} {\bibfnamefont {G.}~\bibnamefont {Ceder}},\ }\bibfield  {title}
  {\bibinfo {title} {Finding nature‚Äôs missing ternary oxide compounds using
  machine learning and density functional theory},\ }\href
  {https://pubs.acs.org/doi/abs/10.1021/cm100795d?casa_token=uudldJSGhRYAAAAA:y9JUSOcCCYhOOgD2A6iIJjqjvbs67dB40m4CMq7dUfe9274lGwlfoBxWLhjY0XJ8kgXUol94g0RdKA}
  {\bibfield  {journal} {\bibinfo  {journal} {Chemistry of Materials}\ }\textbf
  {\bibinfo {volume} {22}},\ \bibinfo {pages} {3762} (\bibinfo {year}
  {2010})}\BibitemShut {NoStop}%
\bibitem [{\citenamefont {Tabor}\ \emph {et~al.}(2018)\citenamefont {Tabor},
  \citenamefont {Roch}, \citenamefont {Saikin}, \citenamefont {Kreisbeck},
  \citenamefont {Sheberla}, \citenamefont {Montoya}, \citenamefont
  {Dwaraknath}, \citenamefont {Aykol}, \citenamefont {Ortiz}, \citenamefont
  {Tribukait} \emph {et~al.}}]{tabor2018accelerating}%
  \BibitemOpen
  \bibfield  {author} {\bibinfo {author} {\bibfnamefont {D.~P.}\ \bibnamefont
  {Tabor}}, \bibinfo {author} {\bibfnamefont {L.~M.}\ \bibnamefont {Roch}},
  \bibinfo {author} {\bibfnamefont {S.~K.}\ \bibnamefont {Saikin}}, \bibinfo
  {author} {\bibfnamefont {C.}~\bibnamefont {Kreisbeck}}, \bibinfo {author}
  {\bibfnamefont {D.}~\bibnamefont {Sheberla}}, \bibinfo {author}
  {\bibfnamefont {J.~H.}\ \bibnamefont {Montoya}}, \bibinfo {author}
  {\bibfnamefont {S.}~\bibnamefont {Dwaraknath}}, \bibinfo {author}
  {\bibfnamefont {M.}~\bibnamefont {Aykol}}, \bibinfo {author} {\bibfnamefont
  {C.}~\bibnamefont {Ortiz}}, \bibinfo {author} {\bibfnamefont
  {H.}~\bibnamefont {Tribukait}}, \emph {et~al.},\ }\bibfield  {title}
  {\bibinfo {title} {Accelerating the discovery of materials for clean energy
  in the era of smart automation},\ }\href
  {https://idp.nature.com/authorize/casa?redirect_uri=https://www.nature.com/articles/s41578-018-0005-z&casa_token=MVKZTbnVt_IAAAAA:11WighMoMi3ITjgovKEIJmp_UbC4Dw-2XkQ39bEDykyUV1kgJcnnu9mT77oVVduHcsEYwEFsypnJZao}
  {\bibfield  {journal} {\bibinfo  {journal} {Nature Reviews Materials}\
  }\textbf {\bibinfo {volume} {3}},\ \bibinfo {pages} {5} (\bibinfo {year}
  {2018})}\BibitemShut {NoStop}%
\bibitem [{\citenamefont {Meredig}\ \emph {et~al.}(2014)\citenamefont
  {Meredig}, \citenamefont {Agrawal}, \citenamefont {Kirklin}, \citenamefont
  {Saal}, \citenamefont {Doak}, \citenamefont {Thompson}, \citenamefont
  {Zhang}, \citenamefont {Choudhary},\ and\ \citenamefont
  {Wolverton}}]{meredig2014combinatorial}%
  \BibitemOpen
  \bibfield  {author} {\bibinfo {author} {\bibfnamefont {B.}~\bibnamefont
  {Meredig}}, \bibinfo {author} {\bibfnamefont {A.}~\bibnamefont {Agrawal}},
  \bibinfo {author} {\bibfnamefont {S.}~\bibnamefont {Kirklin}}, \bibinfo
  {author} {\bibfnamefont {J.~E.}\ \bibnamefont {Saal}}, \bibinfo {author}
  {\bibfnamefont {J.}~\bibnamefont {Doak}}, \bibinfo {author} {\bibfnamefont
  {A.}~\bibnamefont {Thompson}}, \bibinfo {author} {\bibfnamefont
  {K.}~\bibnamefont {Zhang}}, \bibinfo {author} {\bibfnamefont
  {A.}~\bibnamefont {Choudhary}},\ and\ \bibinfo {author} {\bibfnamefont
  {C.}~\bibnamefont {Wolverton}},\ }\bibfield  {title} {\bibinfo {title}
  {Combinatorial screening for new materials in unconstrained composition space
  with machine learning},\ }\href
  {https://journals.aps.org/prb/abstract/10.1103/PhysRevB.89.094104} {\bibfield
   {journal} {\bibinfo  {journal} {Physical Review B}\ }\textbf {\bibinfo
  {volume} {89}},\ \bibinfo {pages} {094104} (\bibinfo {year}
  {2014})}\BibitemShut {NoStop}%
\bibitem [{\citenamefont {Balachandran}\ \emph {et~al.}(2018)\citenamefont
  {Balachandran}, \citenamefont {Emery}, \citenamefont {Gubernatis},
  \citenamefont {Lookman}, \citenamefont {Wolverton},\ and\ \citenamefont
  {Zunger}}]{balachandran2018predictions}%
  \BibitemOpen
  \bibfield  {author} {\bibinfo {author} {\bibfnamefont {P.~V.}\ \bibnamefont
  {Balachandran}}, \bibinfo {author} {\bibfnamefont {A.~A.}\ \bibnamefont
  {Emery}}, \bibinfo {author} {\bibfnamefont {J.~E.}\ \bibnamefont
  {Gubernatis}}, \bibinfo {author} {\bibfnamefont {T.}~\bibnamefont {Lookman}},
  \bibinfo {author} {\bibfnamefont {C.}~\bibnamefont {Wolverton}},\ and\
  \bibinfo {author} {\bibfnamefont {A.}~\bibnamefont {Zunger}},\ }\bibfield
  {title} {\bibinfo {title} {Predictions of new ab o 3 perovskite compounds by
  combining machine learning and density functional theory},\ }\href
  {https://journals.aps.org/prmaterials/abstract/10.1103/PhysRevMaterials.2.043802}
  {\bibfield  {journal} {\bibinfo  {journal} {Physical Review Materials}\
  }\textbf {\bibinfo {volume} {2}},\ \bibinfo {pages} {043802} (\bibinfo {year}
  {2018})}\BibitemShut {NoStop}%
\bibitem [{\citenamefont {Ramprasad}\ \emph {et~al.}(2017)\citenamefont
  {Ramprasad}, \citenamefont {Batra}, \citenamefont {Pilania}, \citenamefont
  {Mannodi-Kanakkithodi},\ and\ \citenamefont {Kim}}]{ramprasad2017machine}%
  \BibitemOpen
  \bibfield  {author} {\bibinfo {author} {\bibfnamefont {R.}~\bibnamefont
  {Ramprasad}}, \bibinfo {author} {\bibfnamefont {R.}~\bibnamefont {Batra}},
  \bibinfo {author} {\bibfnamefont {G.}~\bibnamefont {Pilania}}, \bibinfo
  {author} {\bibfnamefont {A.}~\bibnamefont {Mannodi-Kanakkithodi}},\ and\
  \bibinfo {author} {\bibfnamefont {C.}~\bibnamefont {Kim}},\ }\bibfield
  {title} {\bibinfo {title} {Machine learning in materials informatics: recent
  applications and prospects},\ }\href
  {https://www.nature.com/articles/s41524-017-0056-5} {\bibfield  {journal}
  {\bibinfo  {journal} {npj Computational Materials}\ }\textbf {\bibinfo
  {volume} {3}},\ \bibinfo {pages} {1} (\bibinfo {year} {2017})}\BibitemShut
  {NoStop}%
\bibitem [{\citenamefont {Ward}\ and\ \citenamefont
  {Wolverton}(2017)}]{ward2017atomistic}%
  \BibitemOpen
  \bibfield  {author} {\bibinfo {author} {\bibfnamefont {L.}~\bibnamefont
  {Ward}}\ and\ \bibinfo {author} {\bibfnamefont {C.}~\bibnamefont
  {Wolverton}},\ }\bibfield  {title} {\bibinfo {title} {Atomistic calculations
  and materials informatics: A review},\ }\href
  {https://www.sciencedirect.com/science/article/pii/S1359028616301085?casa_token=iF5ST_jX5ioAAAAA:Wv35MpSu_wZfuCj-ygCv820DBP4hzNbW66AEJW47ESRnIhmzhI4759XSABKr2iCp7VVUwY1s}
  {\bibfield  {journal} {\bibinfo  {journal} {Current Opinion in Solid State
  and Materials Science}\ }\textbf {\bibinfo {volume} {21}},\ \bibinfo {pages}
  {167} (\bibinfo {year} {2017})}\BibitemShut {NoStop}%
\bibitem [{\citenamefont {Ren}\ \emph {et~al.}(2018)\citenamefont {Ren},
  \citenamefont {Ward}, \citenamefont {Williams}, \citenamefont {Laws},
  \citenamefont {Wolverton}, \citenamefont {Hattrick-Simpers},\ and\
  \citenamefont {Mehta}}]{ren2018accelerated}%
  \BibitemOpen
  \bibfield  {author} {\bibinfo {author} {\bibfnamefont {F.}~\bibnamefont
  {Ren}}, \bibinfo {author} {\bibfnamefont {L.}~\bibnamefont {Ward}}, \bibinfo
  {author} {\bibfnamefont {T.}~\bibnamefont {Williams}}, \bibinfo {author}
  {\bibfnamefont {K.~J.}\ \bibnamefont {Laws}}, \bibinfo {author}
  {\bibfnamefont {C.}~\bibnamefont {Wolverton}}, \bibinfo {author}
  {\bibfnamefont {J.}~\bibnamefont {Hattrick-Simpers}},\ and\ \bibinfo {author}
  {\bibfnamefont {A.}~\bibnamefont {Mehta}},\ }\bibfield  {title} {\bibinfo
  {title} {Accelerated discovery of metallic glasses through iteration of
  machine learning and high-throughput experiments},\ }\href
  {https://advances.sciencemag.org/content/4/4/eaaq1566?intcmp\u003dtrendmd-adv}
  {\bibfield  {journal} {\bibinfo  {journal} {Science Advances}\ }\textbf
  {\bibinfo {volume} {4}},\ \bibinfo {pages} {eaaq1566} (\bibinfo {year}
  {2018})}\BibitemShut {NoStop}%
\bibitem [{\citenamefont {Kim}\ \emph {et~al.}(2018)\citenamefont {Kim},
  \citenamefont {Ward}, \citenamefont {He}, \citenamefont {Krishna},
  \citenamefont {Agrawal},\ and\ \citenamefont {Wolverton}}]{kim2018machine}%
  \BibitemOpen
  \bibfield  {author} {\bibinfo {author} {\bibfnamefont {K.}~\bibnamefont
  {Kim}}, \bibinfo {author} {\bibfnamefont {L.}~\bibnamefont {Ward}}, \bibinfo
  {author} {\bibfnamefont {J.}~\bibnamefont {He}}, \bibinfo {author}
  {\bibfnamefont {A.}~\bibnamefont {Krishna}}, \bibinfo {author} {\bibfnamefont
  {A.}~\bibnamefont {Agrawal}},\ and\ \bibinfo {author} {\bibfnamefont
  {C.}~\bibnamefont {Wolverton}},\ }\bibfield  {title} {\bibinfo {title}
  {Machine-learning-accelerated high-throughput materials screening: Discovery
  of novel quaternary heusler compounds},\ }\href
  {https://journals.aps.org/prmaterials/abstract/10.1103/PhysRevMaterials.2.123801}
  {\bibfield  {journal} {\bibinfo  {journal} {Physical Review Materials}\
  }\textbf {\bibinfo {volume} {2}},\ \bibinfo {pages} {123801} (\bibinfo {year}
  {2018})}\BibitemShut {NoStop}%
\bibitem [{\citenamefont {Gautier}\ \emph {et~al.}(2015)\citenamefont
  {Gautier}, \citenamefont {Zhang}, \citenamefont {Hu}, \citenamefont {Yu},
  \citenamefont {Lin}, \citenamefont {Sunde}, \citenamefont {Chon},
  \citenamefont {Poeppelmeier},\ and\ \citenamefont
  {Zunger}}]{gautier2015prediction}%
  \BibitemOpen
  \bibfield  {author} {\bibinfo {author} {\bibfnamefont {R.}~\bibnamefont
  {Gautier}}, \bibinfo {author} {\bibfnamefont {X.}~\bibnamefont {Zhang}},
  \bibinfo {author} {\bibfnamefont {L.}~\bibnamefont {Hu}}, \bibinfo {author}
  {\bibfnamefont {L.}~\bibnamefont {Yu}}, \bibinfo {author} {\bibfnamefont
  {Y.}~\bibnamefont {Lin}}, \bibinfo {author} {\bibfnamefont {T.~O.}\
  \bibnamefont {Sunde}}, \bibinfo {author} {\bibfnamefont {D.}~\bibnamefont
  {Chon}}, \bibinfo {author} {\bibfnamefont {K.~R.}\ \bibnamefont
  {Poeppelmeier}},\ and\ \bibinfo {author} {\bibfnamefont {A.}~\bibnamefont
  {Zunger}},\ }\bibfield  {title} {\bibinfo {title} {Prediction and accelerated
  laboratory discovery of previously unknown 18-electron abx compounds},\
  }\href {https://www.nature.com/articles/nchem.2207.pdf?origin=ppub}
  {\bibfield  {journal} {\bibinfo  {journal} {Nature chemistry}\ }\textbf
  {\bibinfo {volume} {7}},\ \bibinfo {pages} {308} (\bibinfo {year}
  {2015})}\BibitemShut {NoStop}%
\bibitem [{\citenamefont {He}\ \emph {et~al.}(2018)\citenamefont {He},
  \citenamefont {Naghavi}, \citenamefont {Hegde}, \citenamefont {Amsler},\ and\
  \citenamefont {Wolverton}}]{he2018designing}%
  \BibitemOpen
  \bibfield  {author} {\bibinfo {author} {\bibfnamefont {J.}~\bibnamefont
  {He}}, \bibinfo {author} {\bibfnamefont {S.~S.}\ \bibnamefont {Naghavi}},
  \bibinfo {author} {\bibfnamefont {V.~I.}\ \bibnamefont {Hegde}}, \bibinfo
  {author} {\bibfnamefont {M.}~\bibnamefont {Amsler}},\ and\ \bibinfo {author}
  {\bibfnamefont {C.}~\bibnamefont {Wolverton}},\ }\bibfield  {title} {\bibinfo
  {title} {Designing and discovering a new family of semiconducting quaternary
  heusler compounds based on the 18-electron rule},\ }\href
  {https://pubs.acs.org/doi/abs/10.1021/acs.chemmater.8b01096?casa_token=Quz-J8hINuIAAAAA:vVusMde4NIyPXquTSdywuxrABMUGTJ_GbsLffKver0Y2K74BvYb0VWM6yyXb64vynSN28TS5HbLZog}
  {\bibfield  {journal} {\bibinfo  {journal} {Chemistry of Materials}\ }\textbf
  {\bibinfo {volume} {30}},\ \bibinfo {pages} {4978} (\bibinfo {year}
  {2018})}\BibitemShut {NoStop}%
\bibitem [{\citenamefont {Koscielski}\ and\ \citenamefont
  {Ibers}(2012)}]{koscielski2012structural}%
  \BibitemOpen
  \bibfield  {author} {\bibinfo {author} {\bibfnamefont {L.~A.}\ \bibnamefont
  {Koscielski}}\ and\ \bibinfo {author} {\bibfnamefont {J.~A.}\ \bibnamefont
  {Ibers}},\ }\bibfield  {title} {\bibinfo {title} {The structural chemistry of
  quaternary chalcogenides of the type amm'q3},\ }\href
  {https://onlinelibrary.wiley.com/doi/full/10.1002/zaac.201200301?casa_token=iglYZJwCs7YAAAAA%3Ac0RxWgDh70NhyoQe_Hk6vqgBgt6ZTN6tCe5-byHqmKlwk7IuvgvNHL1sfF4ejsJo9vBMTBRItJJv}
  {\bibfield  {journal} {\bibinfo  {journal} {Zeitschrift f{\"u}r anorganische
  und allgemeine Chemie}\ }\textbf {\bibinfo {volume} {638}},\ \bibinfo {pages}
  {2585} (\bibinfo {year} {2012})}\BibitemShut {NoStop}%
\bibitem [{\citenamefont {Strobel}\ and\ \citenamefont
  {Schleid}(2006)}]{strobel2006three}%
  \BibitemOpen
  \bibfield  {author} {\bibinfo {author} {\bibfnamefont {S.}~\bibnamefont
  {Strobel}}\ and\ \bibinfo {author} {\bibfnamefont {T.}~\bibnamefont
  {Schleid}},\ }\bibfield  {title} {\bibinfo {title} {Three structure types for
  strontium copper (i) lanthanide (iii) selenides srcumse3 (m= la, gd, lu)},\
  }\href
  {https://www.sciencedirect.com/science/article/pii/S0925838805017937?casa_token=HzPM2qAqfHsAAAAA:jk-8xMR5IeIBDMPbymnsbeBhsRyxcyKQo9tXRnQhnSp0dpYTXa9tyAH6NkVLWp858M8QAQw}
  {\bibfield  {journal} {\bibinfo  {journal} {Journal of alloys and compounds}\
  }\textbf {\bibinfo {volume} {418}},\ \bibinfo {pages} {80} (\bibinfo {year}
  {2006})}\BibitemShut {NoStop}%
\bibitem [{\citenamefont {Ruseikina}\ \emph {et~al.}(2019)\citenamefont
  {Ruseikina}, \citenamefont {Solovyov}, \citenamefont {Chernyshev},
  \citenamefont {Aleksandrovsky}, \citenamefont {Andreev}, \citenamefont
  {Krylova}, \citenamefont {Krylov}, \citenamefont {Velikanov}, \citenamefont
  {Molokeev}, \citenamefont {Maximov} \emph {et~al.}}]{ruseikina2019synthesis}%
  \BibitemOpen
  \bibfield  {author} {\bibinfo {author} {\bibfnamefont {A.~V.}\ \bibnamefont
  {Ruseikina}}, \bibinfo {author} {\bibfnamefont {L.~A.}\ \bibnamefont
  {Solovyov}}, \bibinfo {author} {\bibfnamefont {V.~–.}\ \bibnamefont
  {Chernyshev}}, \bibinfo {author} {\bibfnamefont {A.~S.}\ \bibnamefont
  {Aleksandrovsky}}, \bibinfo {author} {\bibfnamefont {O.~V.}\ \bibnamefont
  {Andreev}}, \bibinfo {author} {\bibfnamefont {S.~N.}\ \bibnamefont
  {Krylova}}, \bibinfo {author} {\bibfnamefont {A.~S.}\ \bibnamefont {Krylov}},
  \bibinfo {author} {\bibfnamefont {D.~–.}\ \bibnamefont {Velikanov}}, \bibinfo
  {author} {\bibfnamefont {M.~S.}\ \bibnamefont {Molokeev}}, \bibinfo {author}
  {\bibfnamefont {N.~G.}\ \bibnamefont {Maximov}}, \emph {et~al.},\ }\bibfield
  {title} {\bibinfo {title} {Synthesis, structure, and properties of
  euercus3},\ }\href
  {https://www.sciencedirect.com/science/article/pii/S0925838819325538?casa_token=LYTYzjK3jDMAAAAA:fNalBqNHW9QXmxnWvsUMq0rLExGeY7jLno5PXI6E6KJTWQGDDntTpwi5EWYOSV9hUZKfo0M}
  {\bibfield  {journal} {\bibinfo  {journal} {Journal of Alloys and Compounds}\
  }\textbf {\bibinfo {volume} {805}},\ \bibinfo {pages} {779} (\bibinfo {year}
  {2019})}\BibitemShut {NoStop}%
\bibitem [{\citenamefont {Maier}\ \emph {et~al.}(2016)\citenamefont {Maier},
  \citenamefont {Prakash}, \citenamefont {Berthebaud}, \citenamefont {Perez},
  \citenamefont {Bobev},\ and\ \citenamefont {Gascoin}}]{maier2016crystal}%
  \BibitemOpen
  \bibfield  {author} {\bibinfo {author} {\bibfnamefont {S.}~\bibnamefont
  {Maier}}, \bibinfo {author} {\bibfnamefont {J.}~\bibnamefont {Prakash}},
  \bibinfo {author} {\bibfnamefont {D.}~\bibnamefont {Berthebaud}}, \bibinfo
  {author} {\bibfnamefont {O.}~\bibnamefont {Perez}}, \bibinfo {author}
  {\bibfnamefont {S.}~\bibnamefont {Bobev}},\ and\ \bibinfo {author}
  {\bibfnamefont {F.}~\bibnamefont {Gascoin}},\ }\bibfield  {title} {\bibinfo
  {title} {Crystal structures of the four new quaternary copper (i)-selenides
  a0. 5cuzrse3 and acuyse3 (a= sr, ba)},\ }\href
  {https://www.sciencedirect.com/science/article/pii/S002245961630250X?casa_token=x3BJ8cwqDmcAAAAA:hnQLZOVvlGYFsMEPZSWnxuFgIs-GTcVjzO8Sj_p6JUSthHnTOGrCGZWVRSZ_q_pRo-hMsWo}
  {\bibfield  {journal} {\bibinfo  {journal} {Journal of Solid State
  Chemistry}\ }\textbf {\bibinfo {volume} {242}},\ \bibinfo {pages} {14}
  (\bibinfo {year} {2016})}\BibitemShut {NoStop}%
\bibitem [{\citenamefont {Ruseikina}\ \emph {et~al.}(2017)\citenamefont
  {Ruseikina}, \citenamefont {Andreev}, \citenamefont {Galenko},\ and\
  \citenamefont {Koltsov}}]{ruseikina2017trends}%
  \BibitemOpen
  \bibfield  {author} {\bibinfo {author} {\bibfnamefont {A.~V.}\ \bibnamefont
  {Ruseikina}}, \bibinfo {author} {\bibfnamefont {O.~V.}\ \bibnamefont
  {Andreev}}, \bibinfo {author} {\bibfnamefont {E.~O.}\ \bibnamefont
  {Galenko}},\ and\ \bibinfo {author} {\bibfnamefont {S.~I.}\ \bibnamefont
  {Koltsov}},\ }\bibfield  {title} {\bibinfo {title} {Trends in thermodynamic
  parameters of phase transitions of lanthanide sulfides srlncus 3 (ln=
  la--lu)},\ }\href
  {https://idp.springer.com/authorize/casa?redirect_uri=https://link.springer.com/content/pdf/10.1007/s10973-016-6010-9.pdf&casa_token=39Hk5dP-8TEAAAAA:jR4kZVE3Xc9F78hgJ80WChvpQ9tNxqAQQN7P3rnmThUbjNS5eMQssntv_XDL7V2lfDTWFyPv0fSZz2Q}
  {\bibfield  {journal} {\bibinfo  {journal} {Journal of Thermal Analysis and
  Calorimetry}\ }\textbf {\bibinfo {volume} {128}},\ \bibinfo {pages} {993}
  (\bibinfo {year} {2017})}\BibitemShut {NoStop}%
\bibitem [{\citenamefont {Ruseikina}\ \emph {et~al.}(2018)\citenamefont
  {Ruseikina}, \citenamefont {Solov‚Äôev}, \citenamefont {Galenko},\ and\
  \citenamefont {Grigor‚Äôev}}]{ruseikina2018refined}%
  \BibitemOpen
  \bibfield  {author} {\bibinfo {author} {\bibfnamefont {A.}~\bibnamefont
  {Ruseikina}}, \bibinfo {author} {\bibfnamefont {L.}~\bibnamefont
  {Solov‚Äôev}}, \bibinfo {author} {\bibfnamefont {E.}~\bibnamefont
  {Galenko}},\ and\ \bibinfo {author} {\bibfnamefont {M.}~\bibnamefont
  {Grigor‚Äôev}},\ }\bibfield  {title} {\bibinfo {title} {Refined crystal
  structures of srlncus 3 (ln= er, yb)},\ }\href
  {https://link.springer.com/article/10.1134/S0036023618090140} {\bibfield
  {journal} {\bibinfo  {journal} {Russian Journal of Inorganic Chemistry}\
  }\textbf {\bibinfo {volume} {63}},\ \bibinfo {pages} {1225} (\bibinfo {year}
  {2018})}\BibitemShut {NoStop}%
\bibitem [{\citenamefont {Sikerina}\ and\ \citenamefont
  {Andreev}(2007)}]{sikerina2007crystal}%
  \BibitemOpen
  \bibfield  {author} {\bibinfo {author} {\bibfnamefont {N.}~\bibnamefont
  {Sikerina}}\ and\ \bibinfo {author} {\bibfnamefont {O.}~\bibnamefont
  {Andreev}},\ }\bibfield  {title} {\bibinfo {title} {Crystal structures of
  srlncus 3 (ln= gd, lu)},\ }\href
  {https://idp.springer.com/authorize/casa?redirect_uri=https://link.springer.com/article/10.1134/S0036023607040183&casa_token=tjUv-ZK4VMgAAAAA:b1r3FOnWYbLPAhbkSOTYCUatV1mOLO7rRf_va9uc01AQjnEesmRJNOEVeyBvXNLF5kTSOe0JLeNhjA}
  {\bibfield  {journal} {\bibinfo  {journal} {Russian Journal of Inorganic
  Chemistry}\ }\textbf {\bibinfo {volume} {52}},\ \bibinfo {pages} {581}
  (\bibinfo {year} {2007})}\BibitemShut {NoStop}%
\bibitem [{\citenamefont {Prakash}\ \emph {et~al.}(2015)\citenamefont
  {Prakash}, \citenamefont {Mesbah}, \citenamefont {Beard},\ and\ \citenamefont
  {Ibers}}]{prakash2015syntheses}%
  \BibitemOpen
  \bibfield  {author} {\bibinfo {author} {\bibfnamefont {J.}~\bibnamefont
  {Prakash}}, \bibinfo {author} {\bibfnamefont {A.}~\bibnamefont {Mesbah}},
  \bibinfo {author} {\bibfnamefont {J.~C.}\ \bibnamefont {Beard}},\ and\
  \bibinfo {author} {\bibfnamefont {J.~A.}\ \bibnamefont {Ibers}},\ }\bibfield
  {title} {\bibinfo {title} {Syntheses and crystal structures of baagtbs3,
  bacugdte3, bacutbte3, baagtbte3, and csagute3},\ }\href
  {https://onlinelibrary.wiley.com/doi/abs/10.1002/zaac.201500027?casa_token=94siRxOSaWEAAAAA:sBOh5GwEAPAGBkgZYSY6UndX0vi7J8O92ZcQQ-x36GXxZ7E_ItbvHTBKWKCGzEFVbJq3tvIUt9n1}
  {\bibfield  {journal} {\bibinfo  {journal} {Zeitschrift f{\"u}r anorganische
  und allgemeine Chemie}\ }\textbf {\bibinfo {volume} {641}},\ \bibinfo {pages}
  {1253} (\bibinfo {year} {2015})}\BibitemShut {NoStop}%
\bibitem [{\citenamefont {Pal}\ \emph {et~al.}(2019{\natexlab{a}})\citenamefont
  {Pal}, \citenamefont {Xia}, \citenamefont {He},\ and\ \citenamefont
  {Wolverton}}]{pal2019intrinsically}%
  \BibitemOpen
  \bibfield  {author} {\bibinfo {author} {\bibfnamefont {K.}~\bibnamefont
  {Pal}}, \bibinfo {author} {\bibfnamefont {Y.}~\bibnamefont {Xia}}, \bibinfo
  {author} {\bibfnamefont {J.}~\bibnamefont {He}},\ and\ \bibinfo {author}
  {\bibfnamefont {C.}~\bibnamefont {Wolverton}},\ }\bibfield  {title} {\bibinfo
  {title} {Intrinsically low lattice thermal conductivity derived from rattler
  cations in an amm'q3 family of chalcogenides},\ }\href
  {https://pubs.acs.org/doi/abs/10.1021/acs.chemmater.9b02484?casa_token=hbemdIHKYX0AAAAA:olaUGE7xQ0JBLBHcZGC0JoIzGFX6QfcwajRELkhEjk04pbA_UjjXXmyL3LyKtX2SCLWAOccCYRJj3w}
  {\bibfield  {journal} {\bibinfo  {journal} {Chem. Mater.}\ }\textbf {\bibinfo
  {volume} {31}},\ \bibinfo {pages} {8734} (\bibinfo {year}
  {2019}{\natexlab{a}})}\BibitemShut {NoStop}%
\bibitem [{\citenamefont {Hao}\ \emph {et~al.}(2019)\citenamefont {Hao},
  \citenamefont {Ward}, \citenamefont {Luo}, \citenamefont {Ozolins},
  \citenamefont {Dravid}, \citenamefont {Kanatzidis},\ and\ \citenamefont
  {Wolverton}}]{hao2019design}%
  \BibitemOpen
  \bibfield  {author} {\bibinfo {author} {\bibfnamefont {S.}~\bibnamefont
  {Hao}}, \bibinfo {author} {\bibfnamefont {L.}~\bibnamefont {Ward}}, \bibinfo
  {author} {\bibfnamefont {Z.}~\bibnamefont {Luo}}, \bibinfo {author}
  {\bibfnamefont {V.}~\bibnamefont {Ozolins}}, \bibinfo {author} {\bibfnamefont
  {V.~P.}\ \bibnamefont {Dravid}}, \bibinfo {author} {\bibfnamefont {M.~G.}\
  \bibnamefont {Kanatzidis}},\ and\ \bibinfo {author} {\bibfnamefont
  {C.}~\bibnamefont {Wolverton}},\ }\bibfield  {title} {\bibinfo {title}
  {Design strategy for high-performance thermoelectric materials: The
  prediction of electron-doped kzrcuse3},\ }\href
  {https://pubs.acs.org/doi/abs/10.1021/acs.chemmater.9b00840?casa_token=toM3_DZLz2EAAAAA:Cc7utT-xNfST9tFgvQelBzstRJN8jkA29z4mm3nZPWPj0rbg5zWtK4Ifaj4krkGCaDF-K0U88nj12g}
  {\bibfield  {journal} {\bibinfo  {journal} {Chemistry of Materials}\ }\textbf
  {\bibinfo {volume} {31}},\ \bibinfo {pages} {3018} (\bibinfo {year}
  {2019})}\BibitemShut {NoStop}%
\bibitem [{\citenamefont {Pal}\ \emph {et~al.}(2019{\natexlab{b}})\citenamefont
  {Pal}, \citenamefont {Xia}, \citenamefont {He},\ and\ \citenamefont
  {Wolverton}}]{pal2019high}%
  \BibitemOpen
  \bibfield  {author} {\bibinfo {author} {\bibfnamefont {K.}~\bibnamefont
  {Pal}}, \bibinfo {author} {\bibfnamefont {Y.}~\bibnamefont {Xia}}, \bibinfo
  {author} {\bibfnamefont {J.}~\bibnamefont {He}},\ and\ \bibinfo {author}
  {\bibfnamefont {C.}~\bibnamefont {Wolverton}},\ }\bibfield  {title} {\bibinfo
  {title} {High thermoelectric performance in baagyte 3 via low lattice thermal
  conductivity induced by bonding heterogeneity},\ }\href
  {https://journals.aps.org/prmaterials/abstract/10.1103/PhysRevMaterials.3.085402}
  {\bibfield  {journal} {\bibinfo  {journal} {Physical Review Materials}\
  }\textbf {\bibinfo {volume} {3}},\ \bibinfo {pages} {085402} (\bibinfo {year}
  {2019}{\natexlab{b}})}\BibitemShut {NoStop}%
\bibitem [{\citenamefont {Pal}\ \emph {et~al.}(2019{\natexlab{c}})\citenamefont
  {Pal}, \citenamefont {Hua}, \citenamefont {Xia},\ and\ \citenamefont
  {Wolverton}}]{pal2019unraveling}%
  \BibitemOpen
  \bibfield  {author} {\bibinfo {author} {\bibfnamefont {K.}~\bibnamefont
  {Pal}}, \bibinfo {author} {\bibfnamefont {X.}~\bibnamefont {Hua}}, \bibinfo
  {author} {\bibfnamefont {Y.}~\bibnamefont {Xia}},\ and\ \bibinfo {author}
  {\bibfnamefont {C.}~\bibnamefont {Wolverton}},\ }\bibfield  {title} {\bibinfo
  {title} {Unraveling the structure-valence-property relationships in amm'q3
  chalcogenides with promising thermoelectric performance},\ }\href
  {https://pubs.acs.org/doi/abs/10.1021/acsaem.9b02139?casa_token=UmEYdgJpGcAAAAAA:jSt8ARmqTmI-LurWzPaMHu3gKE24oGwna0MMwiXy9jGE7qLxGYYUnoHs9Sfd3WM4oVaqa0mPALXPXQ}
  {\bibfield  {journal} {\bibinfo  {journal} {ACS Applied Energy Materials}\
  }\textbf {\bibinfo {volume} {3}},\ \bibinfo {pages} {2110} (\bibinfo {year}
  {2019}{\natexlab{c}})}\BibitemShut {NoStop}%
\bibitem [{\citenamefont {Fabini}\ \emph {et~al.}(2019)\citenamefont {Fabini},
  \citenamefont {Koerner},\ and\ \citenamefont
  {Seshadri}}]{fabini2019candidate}%
  \BibitemOpen
  \bibfield  {author} {\bibinfo {author} {\bibfnamefont {D.~H.}\ \bibnamefont
  {Fabini}}, \bibinfo {author} {\bibfnamefont {M.}~\bibnamefont {Koerner}},\
  and\ \bibinfo {author} {\bibfnamefont {R.}~\bibnamefont {Seshadri}},\
  }\bibfield  {title} {\bibinfo {title} {Candidate inorganic photovoltaic
  materials from electronic structure-based optical absorption and charge
  transport proxies},\ }\href
  {https://pubs.acs.org/doi/abs/10.1021/acs.chemmater.8b04542?casa_token=OAsKPb6h-AEAAAAA:dLKpo-W-evyzxj1bOVfPO6NWLLa8Xv2LhNqb56XMX-_psl4uZP2Zwui8dY13Ne1nUwRkYdCSskeGOw}
  {\bibfield  {journal} {\bibinfo  {journal} {Chemistry of Materials}\ }\textbf
  {\bibinfo {volume} {31}},\ \bibinfo {pages} {1561} (\bibinfo {year}
  {2019})}\BibitemShut {NoStop}%
\bibitem [{\citenamefont {Pal}\ \emph {et~al.}(2021{\natexlab{a}})\citenamefont
  {Pal}, \citenamefont {Xia}, \citenamefont {Shen}, \citenamefont {He},
  \citenamefont {Luo}, \citenamefont {Kanatzidis},\ and\ \citenamefont
  {Wolverton}}]{pal2021accelerated}%
  \BibitemOpen
  \bibfield  {author} {\bibinfo {author} {\bibfnamefont {K.}~\bibnamefont
  {Pal}}, \bibinfo {author} {\bibfnamefont {Y.}~\bibnamefont {Xia}}, \bibinfo
  {author} {\bibfnamefont {J.}~\bibnamefont {Shen}}, \bibinfo {author}
  {\bibfnamefont {J.}~\bibnamefont {He}}, \bibinfo {author} {\bibfnamefont
  {Y.}~\bibnamefont {Luo}}, \bibinfo {author} {\bibfnamefont {M.~G.}\
  \bibnamefont {Kanatzidis}},\ and\ \bibinfo {author} {\bibfnamefont
  {C.}~\bibnamefont {Wolverton}},\ }\bibfield  {title} {\bibinfo {title}
  {Accelerated discovery of a large family of quaternary chalcogenides with
  very low lattice thermal conductivity},\ }\href
  {https://www.nature.com/articles/s41524-021-00549-x} {\bibfield  {journal}
  {\bibinfo  {journal} {npj Computational Materials}\ }\textbf {\bibinfo
  {volume} {7}},\ \bibinfo {pages} {1} (\bibinfo {year}
  {2021}{\natexlab{a}})}\BibitemShut {NoStop}%
\bibitem [{\citenamefont {Park}\ and\ \citenamefont
  {Wolverton}(2020)}]{park2020developing}%
  \BibitemOpen
  \bibfield  {author} {\bibinfo {author} {\bibfnamefont {C.~W.}\ \bibnamefont
  {Park}}\ and\ \bibinfo {author} {\bibfnamefont {C.}~\bibnamefont
  {Wolverton}},\ }\bibfield  {title} {\bibinfo {title} {Developing an improved
  crystal graph convolutional neural network framework for accelerated
  materials discovery},\ }\href
  {https://journals.aps.org/prmaterials/abstract/10.1103/PhysRevMaterials.4.063801}
  {\bibfield  {journal} {\bibinfo  {journal} {Physical Review Materials}\
  }\textbf {\bibinfo {volume} {4}},\ \bibinfo {pages} {063801} (\bibinfo {year}
  {2020})}\BibitemShut {NoStop}%
\bibitem [{\citenamefont {Xie}\ and\ \citenamefont
  {Grossman}(2018)}]{xie2018crystal}%
  \BibitemOpen
  \bibfield  {author} {\bibinfo {author} {\bibfnamefont {T.}~\bibnamefont
  {Xie}}\ and\ \bibinfo {author} {\bibfnamefont {J.~C.}\ \bibnamefont
  {Grossman}},\ }\bibfield  {title} {\bibinfo {title} {Crystal graph
  convolutional neural networks for an accurate and interpretable prediction of
  material properties},\ }\href
  {https://journals.aps.org/prl/abstract/10.1103/PhysRevLett.120.145301}
  {\bibfield  {journal} {\bibinfo  {journal} {Physical Review Letters}\
  }\textbf {\bibinfo {volume} {120}},\ \bibinfo {pages} {145301} (\bibinfo
  {year} {2018})}\BibitemShut {NoStop}%
\bibitem [{\citenamefont {Ward}\ \emph {et~al.}(2017)\citenamefont {Ward},
  \citenamefont {Liu}, \citenamefont {Krishna}, \citenamefont {Hegde},
  \citenamefont {Agrawal}, \citenamefont {Choudhary},\ and\ \citenamefont
  {Wolverton}}]{ward2017including}%
  \BibitemOpen
  \bibfield  {author} {\bibinfo {author} {\bibfnamefont {L.}~\bibnamefont
  {Ward}}, \bibinfo {author} {\bibfnamefont {R.}~\bibnamefont {Liu}}, \bibinfo
  {author} {\bibfnamefont {A.}~\bibnamefont {Krishna}}, \bibinfo {author}
  {\bibfnamefont {V.~I.}\ \bibnamefont {Hegde}}, \bibinfo {author}
  {\bibfnamefont {A.}~\bibnamefont {Agrawal}}, \bibinfo {author} {\bibfnamefont
  {A.}~\bibnamefont {Choudhary}},\ and\ \bibinfo {author} {\bibfnamefont
  {C.}~\bibnamefont {Wolverton}},\ }\bibfield  {title} {\bibinfo {title}
  {Including crystal structure attributes in machine learning models of
  formation energies via voronoi tessellations},\ }\href@noop {} {\bibfield
  {journal} {\bibinfo  {journal} {Physical Review B}\ }\textbf {\bibinfo
  {volume} {96}},\ \bibinfo {pages} {024104} (\bibinfo {year}
  {2017})}\BibitemShut {NoStop}%
\bibitem [{\citenamefont {Wu}\ \emph {et~al.}(2017)\citenamefont {Wu},
  \citenamefont {Wu}, \citenamefont {He}, \citenamefont {Zhao}, \citenamefont
  {Li}, \citenamefont {Wu}, \citenamefont {Jin}, \citenamefont {Xu},
  \citenamefont {Jiang}, \citenamefont {Huang}, \citenamefont {Zhu},\ and\
  \citenamefont {He}}]{wu2017direct}%
  \BibitemOpen
  \bibfield  {author} {\bibinfo {author} {\bibfnamefont {D.}~\bibnamefont
  {Wu}}, \bibinfo {author} {\bibfnamefont {L.}~\bibnamefont {Wu}}, \bibinfo
  {author} {\bibfnamefont {D.}~\bibnamefont {He}}, \bibinfo {author}
  {\bibfnamefont {L.-D.}\ \bibnamefont {Zhao}}, \bibinfo {author}
  {\bibfnamefont {W.}~\bibnamefont {Li}}, \bibinfo {author} {\bibfnamefont
  {M.}~\bibnamefont {Wu}}, \bibinfo {author} {\bibfnamefont {M.}~\bibnamefont
  {Jin}}, \bibinfo {author} {\bibfnamefont {J.}~\bibnamefont {Xu}}, \bibinfo
  {author} {\bibfnamefont {J.}~\bibnamefont {Jiang}}, \bibinfo {author}
  {\bibfnamefont {L.}~\bibnamefont {Huang}}, \bibinfo {author} {\bibfnamefont
  {M.~G.}\ \bibnamefont {Zhu}, \bibfnamefont {Yimei~Kanatzidis}},\ and\
  \bibinfo {author} {\bibfnamefont {J.}~\bibnamefont {He}},\ }\bibfield
  {title} {\bibinfo {title} {Direct observation of vast off-stoichiometric
  defects in single crystalline snse},\ }\href
  {https://www.sciencedirect.com/science/article/pii/S2211285517302082?casa_token=oLj8TaJ27y4AAAAA:IDwCy_8Cg-mky-zQGSKc1l3w_QrrPJEMJlTl0k2P2rAzQHE8GIhqIAfnLjZcvAm6JInviirT}
  {\bibfield  {journal} {\bibinfo  {journal} {Nano Energy}\ }\textbf {\bibinfo
  {volume} {35}},\ \bibinfo {pages} {321} (\bibinfo {year} {2017})}\BibitemShut
  {NoStop}%
\bibitem [{\citenamefont {Morelli}\ \emph {et~al.}(2008)\citenamefont
  {Morelli}, \citenamefont {Jovovic},\ and\ \citenamefont
  {Heremans}}]{morelli2008intrinsically}%
  \BibitemOpen
  \bibfield  {author} {\bibinfo {author} {\bibfnamefont {D.}~\bibnamefont
  {Morelli}}, \bibinfo {author} {\bibfnamefont {V.}~\bibnamefont {Jovovic}},\
  and\ \bibinfo {author} {\bibfnamefont {J.}~\bibnamefont {Heremans}},\
  }\bibfield  {title} {\bibinfo {title} {Intrinsically minimal thermal
  conductivity in cubic i- v- vi 2 semiconductors},\ }\href@noop {} {\bibfield
  {journal} {\bibinfo  {journal} {Phys. Rev. Lett.}\ }\textbf {\bibinfo
  {volume} {101}},\ \bibinfo {pages} {035901} (\bibinfo {year}
  {2008})}\BibitemShut {NoStop}%
\bibitem [{\citenamefont {Xia}\ \emph {et~al.}(2020{\natexlab{b}})\citenamefont
  {Xia}, \citenamefont {Hegde}, \citenamefont {Pal}, \citenamefont {Hua},
  \citenamefont {Gaines}, \citenamefont {Patel}, \citenamefont {He},
  \citenamefont {Aykol},\ and\ \citenamefont {Wolverton}}]{xia2020high}%
  \BibitemOpen
  \bibfield  {author} {\bibinfo {author} {\bibfnamefont {Y.}~\bibnamefont
  {Xia}}, \bibinfo {author} {\bibfnamefont {V.~I.}\ \bibnamefont {Hegde}},
  \bibinfo {author} {\bibfnamefont {K.}~\bibnamefont {Pal}}, \bibinfo {author}
  {\bibfnamefont {X.}~\bibnamefont {Hua}}, \bibinfo {author} {\bibfnamefont
  {D.}~\bibnamefont {Gaines}}, \bibinfo {author} {\bibfnamefont
  {S.}~\bibnamefont {Patel}}, \bibinfo {author} {\bibfnamefont
  {J.}~\bibnamefont {He}}, \bibinfo {author} {\bibfnamefont {M.}~\bibnamefont
  {Aykol}},\ and\ \bibinfo {author} {\bibfnamefont {C.}~\bibnamefont
  {Wolverton}},\ }\bibfield  {title} {\bibinfo {title} {High-throughput study
  of lattice thermal conductivity in binary rocksalt and zinc blende compounds
  including higher-order anharmonicity},\ }\href
  {https://journals.aps.org/prx/abstract/10.1103/PhysRevX.10.041029} {\bibfield
   {journal} {\bibinfo  {journal} {Physical Review X}\ }\textbf {\bibinfo
  {volume} {10}},\ \bibinfo {pages} {041029} (\bibinfo {year}
  {2020}{\natexlab{b}})}\BibitemShut {NoStop}%
\bibitem [{\citenamefont {Pal}\ \emph {et~al.}(2021{\natexlab{b}})\citenamefont
  {Pal}, \citenamefont {Xia},\ and\ \citenamefont
  {Wolverton}}]{pal2021microscopic}%
  \BibitemOpen
  \bibfield  {author} {\bibinfo {author} {\bibfnamefont {K.}~\bibnamefont
  {Pal}}, \bibinfo {author} {\bibfnamefont {Y.}~\bibnamefont {Xia}},\ and\
  \bibinfo {author} {\bibfnamefont {C.}~\bibnamefont {Wolverton}},\ }\bibfield
  {title} {\bibinfo {title} {Microscopic mechanism of unusual lattice thermal
  transport in tlinte 2},\ }\href
  {https://www.nature.com/articles/s41524-020-00474-5} {\bibfield  {journal}
  {\bibinfo  {journal} {npj Computational Materials}\ }\textbf {\bibinfo
  {volume} {7}},\ \bibinfo {pages} {1} (\bibinfo {year}
  {2021}{\natexlab{b}})}\BibitemShut {NoStop}%
\bibitem [{\citenamefont {Paszke}\ \emph {et~al.}(2019)\citenamefont {Paszke},
  \citenamefont {Gross}, \citenamefont {Massa}, \citenamefont {Lerer},
  \citenamefont {Bradbury}, \citenamefont {Chanan}, \citenamefont {Killeen},
  \citenamefont {Lin}, \citenamefont {Gimelshein}, \citenamefont {Antiga} \emph
  {et~al.}}]{paszke2019pytorch}%
  \BibitemOpen
  \bibfield  {author} {\bibinfo {author} {\bibfnamefont {A.}~\bibnamefont
  {Paszke}}, \bibinfo {author} {\bibfnamefont {S.}~\bibnamefont {Gross}},
  \bibinfo {author} {\bibfnamefont {F.}~\bibnamefont {Massa}}, \bibinfo
  {author} {\bibfnamefont {A.}~\bibnamefont {Lerer}}, \bibinfo {author}
  {\bibfnamefont {J.}~\bibnamefont {Bradbury}}, \bibinfo {author}
  {\bibfnamefont {G.}~\bibnamefont {Chanan}}, \bibinfo {author} {\bibfnamefont
  {T.}~\bibnamefont {Killeen}}, \bibinfo {author} {\bibfnamefont
  {Z.}~\bibnamefont {Lin}}, \bibinfo {author} {\bibfnamefont {N.}~\bibnamefont
  {Gimelshein}}, \bibinfo {author} {\bibfnamefont {L.}~\bibnamefont {Antiga}},
  \emph {et~al.},\ }\bibfield  {title} {\bibinfo {title} {Pytorch: An
  imperative style, high-performance deep learning library},\ }in\ \href
  {https://papers.nips.cc/paper/2019/hash/bdbca288fee7f92f2bfa9f7012727740-Abstract.html}
  {\emph {\bibinfo {booktitle} {Advances in Neural Information Processing
  Systems}}}\ (\bibinfo {year} {2019})\ pp.\ \bibinfo {pages}
  {8024--8035}\BibitemShut {NoStop}%
\bibitem [{\citenamefont {Ong}\ \emph {et~al.}(2013)\citenamefont {Ong},
  \citenamefont {Richards}, \citenamefont {Jain}, \citenamefont {Hautier},
  \citenamefont {Kocher}, \citenamefont {Cholia}, \citenamefont {Gunter},
  \citenamefont {Chevrier}, \citenamefont {Persson},\ and\ \citenamefont
  {Ceder}}]{ong2013python}%
  \BibitemOpen
  \bibfield  {author} {\bibinfo {author} {\bibfnamefont {S.~P.}\ \bibnamefont
  {Ong}}, \bibinfo {author} {\bibfnamefont {W.~D.}\ \bibnamefont {Richards}},
  \bibinfo {author} {\bibfnamefont {A.}~\bibnamefont {Jain}}, \bibinfo {author}
  {\bibfnamefont {G.}~\bibnamefont {Hautier}}, \bibinfo {author} {\bibfnamefont
  {M.}~\bibnamefont {Kocher}}, \bibinfo {author} {\bibfnamefont
  {S.}~\bibnamefont {Cholia}}, \bibinfo {author} {\bibfnamefont
  {D.}~\bibnamefont {Gunter}}, \bibinfo {author} {\bibfnamefont {V.~L.}\
  \bibnamefont {Chevrier}}, \bibinfo {author} {\bibfnamefont {K.~A.}\
  \bibnamefont {Persson}},\ and\ \bibinfo {author} {\bibfnamefont
  {G.}~\bibnamefont {Ceder}},\ }\bibfield  {title} {\bibinfo {title} {Python
  materials genomics (pymatgen): A robust, open-source python library for
  materials analysis},\ }\href
  {https://www.sciencedirect.com/science/article/pii/S0927025612006295}
  {\bibfield  {journal} {\bibinfo  {journal} {Computational Materials Science}\
  }\textbf {\bibinfo {volume} {68}},\ \bibinfo {pages} {314} (\bibinfo {year}
  {2013})}\BibitemShut {NoStop}%
\bibitem [{\citenamefont {Belsky}\ \emph {et~al.}(2002)\citenamefont {Belsky},
  \citenamefont {Hellenbrandt}, \citenamefont {Karen},\ and\ \citenamefont
  {Luksch}}]{belsky2002new}%
  \BibitemOpen
  \bibfield  {author} {\bibinfo {author} {\bibfnamefont {A.}~\bibnamefont
  {Belsky}}, \bibinfo {author} {\bibfnamefont {M.}~\bibnamefont
  {Hellenbrandt}}, \bibinfo {author} {\bibfnamefont {V.~L.}\ \bibnamefont
  {Karen}},\ and\ \bibinfo {author} {\bibfnamefont {P.}~\bibnamefont
  {Luksch}},\ }\bibfield  {title} {\bibinfo {title} {New developments in the
  inorganic crystal structure database (icsd): accessibility in support of
  materials research and design},\ }\href
  {https://scripts.iucr.org/cgi-bin/paper?an0615} {\bibfield  {journal}
  {\bibinfo  {journal} {Acta Crystallographica Section B: Structural Science}\
  }\textbf {\bibinfo {volume} {58}},\ \bibinfo {pages} {364} (\bibinfo {year}
  {2002})}\BibitemShut {NoStop}%
\bibitem [{\citenamefont {Kresse}\ and\ \citenamefont
  {Furthm{\"u}ller}(1996)}]{kresse1996efficiency}%
  \BibitemOpen
  \bibfield  {author} {\bibinfo {author} {\bibfnamefont {G.}~\bibnamefont
  {Kresse}}\ and\ \bibinfo {author} {\bibfnamefont {J.}~\bibnamefont
  {Furthm{\"u}ller}},\ }\bibfield  {title} {\bibinfo {title} {Efficiency of
  ab-initio total energy calculations for metals and semiconductors using a
  plane-wave basis set},\ }\href
  {https://www.sciencedirect.com/science/article/abs/pii/0927025696000080}
  {\bibfield  {journal} {\bibinfo  {journal} {Comp. Mater. Sci.}\ }\textbf
  {\bibinfo {volume} {6}},\ \bibinfo {pages} {15} (\bibinfo {year}
  {1996})}\BibitemShut {NoStop}%
\bibitem [{\citenamefont {Bl{\"o}chl}(1994)}]{blochl1994projector}%
  \BibitemOpen
  \bibfield  {author} {\bibinfo {author} {\bibfnamefont {P.~E.}\ \bibnamefont
  {Bl{\"o}chl}},\ }\bibfield  {title} {\bibinfo {title} {Projector
  augmented-wave method},\ }\href
  {https://journals.aps.org/prb/abstract/10.1103/PhysRevB.50.17953} {\bibfield
  {journal} {\bibinfo  {journal} {Phys. Rev. B}\ }\textbf {\bibinfo {volume}
  {50}},\ \bibinfo {pages} {17953} (\bibinfo {year} {1994})}\BibitemShut
  {NoStop}%
\bibitem [{\citenamefont {Kresse}\ and\ \citenamefont
  {Joubert}(1999)}]{kresse1999ultrasoft}%
  \BibitemOpen
  \bibfield  {author} {\bibinfo {author} {\bibfnamefont {G.}~\bibnamefont
  {Kresse}}\ and\ \bibinfo {author} {\bibfnamefont {D.}~\bibnamefont
  {Joubert}},\ }\bibfield  {title} {\bibinfo {title} {From ultrasoft
  pseudopotentials to the projector augmented-wave method},\ }\href
  {https://journals.aps.org/prb/abstract/10.1103/PhysRevB.59.1758} {\bibfield
  {journal} {\bibinfo  {journal} {Phys. Rev. B}\ }\textbf {\bibinfo {volume}
  {59}},\ \bibinfo {pages} {1758} (\bibinfo {year} {1999})}\BibitemShut
  {NoStop}%
\bibitem [{\citenamefont {Perdew}\ \emph {et~al.}(1996)\citenamefont {Perdew},
  \citenamefont {Burke},\ and\ \citenamefont
  {Ernzerhof}}]{perdew1996generalized}%
  \BibitemOpen
  \bibfield  {author} {\bibinfo {author} {\bibfnamefont {J.~P.}\ \bibnamefont
  {Perdew}}, \bibinfo {author} {\bibfnamefont {K.}~\bibnamefont {Burke}},\ and\
  \bibinfo {author} {\bibfnamefont {M.}~\bibnamefont {Ernzerhof}},\ }\bibfield
  {title} {\bibinfo {title} {Generalized gradient approximation made simple},\
  }\href {https://journals.aps.org/prl/abstract/10.1103/PhysRevLett.77.3865}
  {\bibfield  {journal} {\bibinfo  {journal} {Phys. Rev. Lett.}\ }\textbf
  {\bibinfo {volume} {77}},\ \bibinfo {pages} {3865} (\bibinfo {year}
  {1996})}\BibitemShut {NoStop}%
\bibitem [{\citenamefont {Zakutayev}\ \emph {et~al.}(2013)\citenamefont
  {Zakutayev}, \citenamefont {Zhang}, \citenamefont {Nagaraja}, \citenamefont
  {Yu}, \citenamefont {Lany}, \citenamefont {Mason}, \citenamefont {Ginley},\
  and\ \citenamefont {Zunger}}]{zakutayev2013theoretical}%
  \BibitemOpen
  \bibfield  {author} {\bibinfo {author} {\bibfnamefont {A.}~\bibnamefont
  {Zakutayev}}, \bibinfo {author} {\bibfnamefont {X.}~\bibnamefont {Zhang}},
  \bibinfo {author} {\bibfnamefont {A.}~\bibnamefont {Nagaraja}}, \bibinfo
  {author} {\bibfnamefont {L.}~\bibnamefont {Yu}}, \bibinfo {author}
  {\bibfnamefont {S.}~\bibnamefont {Lany}}, \bibinfo {author} {\bibfnamefont
  {T.~O.}\ \bibnamefont {Mason}}, \bibinfo {author} {\bibfnamefont {D.~S.}\
  \bibnamefont {Ginley}},\ and\ \bibinfo {author} {\bibfnamefont
  {A.}~\bibnamefont {Zunger}},\ }\bibfield  {title} {\bibinfo {title}
  {Theoretical prediction and experimental realization of new stable inorganic
  materials using the inverse design approach},\ }\href
  {https://pubs.acs.org/doi/abs/10.1021/ja311599g} {\bibfield  {journal}
  {\bibinfo  {journal} {Journal of the American Chemical Society}\ }\textbf
  {\bibinfo {volume} {135}},\ \bibinfo {pages} {10048} (\bibinfo {year}
  {2013})}\BibitemShut {NoStop}%
\bibitem [{\citenamefont {Aykol}\ \emph {et~al.}(2018)\citenamefont {Aykol},
  \citenamefont {Dwaraknath}, \citenamefont {Sun},\ and\ \citenamefont
  {Persson}}]{aykol2018thermodynamic}%
  \BibitemOpen
  \bibfield  {author} {\bibinfo {author} {\bibfnamefont {M.}~\bibnamefont
  {Aykol}}, \bibinfo {author} {\bibfnamefont {S.~S.}\ \bibnamefont
  {Dwaraknath}}, \bibinfo {author} {\bibfnamefont {W.}~\bibnamefont {Sun}},\
  and\ \bibinfo {author} {\bibfnamefont {K.~A.}\ \bibnamefont {Persson}},\
  }\bibfield  {title} {\bibinfo {title} {Thermodynamic limit for synthesis of
  metastable inorganic materials},\ }\href
  {https://advances.sciencemag.org/content/4/4/eaaq0148?intcmp=trendmd-adv}
  {\bibfield  {journal} {\bibinfo  {journal} {Science advances}\ }\textbf
  {\bibinfo {volume} {4}},\ \bibinfo {pages} {eaaq0148} (\bibinfo {year}
  {2018})}\BibitemShut {NoStop}%
\bibitem [{\citenamefont {Anand}\ \emph {et~al.}(2019)\citenamefont {Anand},
  \citenamefont {Wood}, \citenamefont {Xia}, \citenamefont {Wolverton},\ and\
  \citenamefont {Snyder}}]{anand2019double}%
  \BibitemOpen
  \bibfield  {author} {\bibinfo {author} {\bibfnamefont {S.}~\bibnamefont
  {Anand}}, \bibinfo {author} {\bibfnamefont {M.}~\bibnamefont {Wood}},
  \bibinfo {author} {\bibfnamefont {Y.}~\bibnamefont {Xia}}, \bibinfo {author}
  {\bibfnamefont {C.}~\bibnamefont {Wolverton}},\ and\ \bibinfo {author}
  {\bibfnamefont {G.~J.}\ \bibnamefont {Snyder}},\ }\bibfield  {title}
  {\bibinfo {title} {Double half-heuslers},\ }\href
  {https://www.sciencedirect.com/science/article/pii/S2542435119301710}
  {\bibfield  {journal} {\bibinfo  {journal} {Joule}\ }\textbf {\bibinfo
  {volume} {3}},\ \bibinfo {pages} {1226} (\bibinfo {year} {2019})}\BibitemShut
  {NoStop}%
\bibitem [{\citenamefont {Hautier}\ \emph {et~al.}(2012)\citenamefont
  {Hautier}, \citenamefont {Jain},\ and\ \citenamefont
  {Ong}}]{hautier2012computer}%
  \BibitemOpen
  \bibfield  {author} {\bibinfo {author} {\bibfnamefont {G.}~\bibnamefont
  {Hautier}}, \bibinfo {author} {\bibfnamefont {A.}~\bibnamefont {Jain}},\ and\
  \bibinfo {author} {\bibfnamefont {S.~P.}\ \bibnamefont {Ong}},\ }\bibfield
  {title} {\bibinfo {title} {From the computer to the laboratory: materials
  discovery and design using first-principles calculations},\ }\href
  {https://idp.springer.com/authorize/casa?redirect_uri=https://link.springer.com/article/10.1007/s10853-012-6424-0&casa_token=VhERo2nnBZIAAAAA:EyPyqjYUDecny4dFJvFpcJjYxm_979Y9IiI6YEbWcWWuu90sHwezQrDHELl6x6kYaBOq8i4xPaIJeYw}
  {\bibfield  {journal} {\bibinfo  {journal} {Journal of Materials Science}\
  }\textbf {\bibinfo {volume} {47}},\ \bibinfo {pages} {7317} (\bibinfo {year}
  {2012})}\BibitemShut {NoStop}%
\bibitem [{\citenamefont {Sun}\ \emph {et~al.}(2017)\citenamefont {Sun},
  \citenamefont {Holder}, \citenamefont {Orva{\~n}anos}, \citenamefont {Arca},
  \citenamefont {Zakutayev}, \citenamefont {Lany},\ and\ \citenamefont
  {Ceder}}]{sun2017thermodynamic}%
  \BibitemOpen
  \bibfield  {author} {\bibinfo {author} {\bibfnamefont {W.}~\bibnamefont
  {Sun}}, \bibinfo {author} {\bibfnamefont {A.}~\bibnamefont {Holder}},
  \bibinfo {author} {\bibfnamefont {B.}~\bibnamefont {Orva{\~n}anos}}, \bibinfo
  {author} {\bibfnamefont {E.}~\bibnamefont {Arca}}, \bibinfo {author}
  {\bibfnamefont {A.}~\bibnamefont {Zakutayev}}, \bibinfo {author}
  {\bibfnamefont {S.}~\bibnamefont {Lany}},\ and\ \bibinfo {author}
  {\bibfnamefont {G.}~\bibnamefont {Ceder}},\ }\bibfield  {title} {\bibinfo
  {title} {Thermodynamic routes to novel metastable nitrogen-rich nitrides},\
  }\href
  {https://pubs.acs.org/doi/abs/10.1021/acs.chemmater.7b02399?casa_token=4FnA3y-L2f4AAAAA:u5MO-d4ljcRjubEaPOZY8Cm4zZTy5OWIkCoc6aGjkohW0L_FvkhS29SnXE5fXzvvRgtFcCMs8YSLgQ}
  {\bibfield  {journal} {\bibinfo  {journal} {Chemistry of Materials}\ }\textbf
  {\bibinfo {volume} {29}},\ \bibinfo {pages} {6936} (\bibinfo {year}
  {2017})}\BibitemShut {NoStop}%
\bibitem [{\citenamefont {Cerqueira}\ \emph {et~al.}(2015)\citenamefont
  {Cerqueira}, \citenamefont {Lin}, \citenamefont {Amsler}, \citenamefont
  {Goedecker}, \citenamefont {Botti},\ and\ \citenamefont
  {Marques}}]{cerqueira2015identification}%
  \BibitemOpen
  \bibfield  {author} {\bibinfo {author} {\bibfnamefont {T.~F.}\ \bibnamefont
  {Cerqueira}}, \bibinfo {author} {\bibfnamefont {S.}~\bibnamefont {Lin}},
  \bibinfo {author} {\bibfnamefont {M.}~\bibnamefont {Amsler}}, \bibinfo
  {author} {\bibfnamefont {S.}~\bibnamefont {Goedecker}}, \bibinfo {author}
  {\bibfnamefont {S.}~\bibnamefont {Botti}},\ and\ \bibinfo {author}
  {\bibfnamefont {M.~A.}\ \bibnamefont {Marques}},\ }\bibfield  {title}
  {\bibinfo {title} {Identification of novel cu, ag, and au ternary oxides from
  global structural prediction},\ }\href
  {https://pubs.acs.org/doi/abs/10.1021/acs.chemmater.5b00716?casa_token=jPNi7Y40x3sAAAAA:PaJOXU0p2Kz2XkGAM8Zw96dx0MA5BQAfkO70qT8zrJIIaaY91s9SakfsIzx5QdqgVdvAo4S01kfmaA}
  {\bibfield  {journal} {\bibinfo  {journal} {Chemistry of Materials}\ }\textbf
  {\bibinfo {volume} {27}},\ \bibinfo {pages} {4562} (\bibinfo {year}
  {2015})}\BibitemShut {NoStop}%
\bibitem [{\citenamefont {Wu}\ \emph {et~al.}(2013)\citenamefont {Wu},
  \citenamefont {Lazic}, \citenamefont {Hautier}, \citenamefont {Persson},\
  and\ \citenamefont {Ceder}}]{wu2013first}%
  \BibitemOpen
  \bibfield  {author} {\bibinfo {author} {\bibfnamefont {Y.}~\bibnamefont
  {Wu}}, \bibinfo {author} {\bibfnamefont {P.}~\bibnamefont {Lazic}}, \bibinfo
  {author} {\bibfnamefont {G.}~\bibnamefont {Hautier}}, \bibinfo {author}
  {\bibfnamefont {K.}~\bibnamefont {Persson}},\ and\ \bibinfo {author}
  {\bibfnamefont {G.}~\bibnamefont {Ceder}},\ }\bibfield  {title} {\bibinfo
  {title} {First principles high throughput screening of oxynitrides for
  water-splitting photocatalysts},\ }\href
  {https://pubs.rsc.org/no/content/articlehtml/2013/ee/c2ee23482c?casa_token=LE4RYcTKnMwAAAAA:AgX-KR6UbZN67zCihWsB1RsHBdnlRzc1N_py85CP54zrQPlqgR4dYpGyPa6db5lwhpOcDEjn73dI}
  {\bibfield  {journal} {\bibinfo  {journal} {Energy \& environmental science}\
  }\textbf {\bibinfo {volume} {6}},\ \bibinfo {pages} {157} (\bibinfo {year}
  {2013})}\BibitemShut {NoStop}%
\bibitem [{\citenamefont {Emery}\ \emph {et~al.}(2016)\citenamefont {Emery},
  \citenamefont {Saal}, \citenamefont {Kirklin}, \citenamefont {Hegde},\ and\
  \citenamefont {Wolverton}}]{emery2016high}%
  \BibitemOpen
  \bibfield  {author} {\bibinfo {author} {\bibfnamefont {A.~A.}\ \bibnamefont
  {Emery}}, \bibinfo {author} {\bibfnamefont {J.~E.}\ \bibnamefont {Saal}},
  \bibinfo {author} {\bibfnamefont {S.}~\bibnamefont {Kirklin}}, \bibinfo
  {author} {\bibfnamefont {V.~I.}\ \bibnamefont {Hegde}},\ and\ \bibinfo
  {author} {\bibfnamefont {C.}~\bibnamefont {Wolverton}},\ }\bibfield  {title}
  {\bibinfo {title} {High-throughput computational screening of perovskites for
  thermochemical water splitting applications},\ }\href
  {https://pubs.acs.org/doi/abs/10.1021/acs.chemmater.6b01182?casa_token=HUBKChhGatwAAAAA:Sv5WKJjM3eLW1Jt3QX4Wxc9L8XGC6jDEXeMPkzC2YXHd3D-AY_HdsIsBy1kFl0fV2M2Acd2x3c0_dQ}
  {\bibfield  {journal} {\bibinfo  {journal} {Chemistry of Materials}\ }\textbf
  {\bibinfo {volume} {28}},\ \bibinfo {pages} {5621} (\bibinfo {year}
  {2016})}\BibitemShut {NoStop}%
\bibitem [{\citenamefont {Sun}\ \emph {et~al.}(2016)\citenamefont {Sun},
  \citenamefont {Dacek}, \citenamefont {Ong}, \citenamefont {Hautier},
  \citenamefont {Jain}, \citenamefont {Richards}, \citenamefont {Gamst},
  \citenamefont {Persson},\ and\ \citenamefont {Ceder}}]{sun2016thermodynamic}%
  \BibitemOpen
  \bibfield  {author} {\bibinfo {author} {\bibfnamefont {W.}~\bibnamefont
  {Sun}}, \bibinfo {author} {\bibfnamefont {S.~T.}\ \bibnamefont {Dacek}},
  \bibinfo {author} {\bibfnamefont {S.~P.}\ \bibnamefont {Ong}}, \bibinfo
  {author} {\bibfnamefont {G.}~\bibnamefont {Hautier}}, \bibinfo {author}
  {\bibfnamefont {A.}~\bibnamefont {Jain}}, \bibinfo {author} {\bibfnamefont
  {W.~D.}\ \bibnamefont {Richards}}, \bibinfo {author} {\bibfnamefont {A.~C.}\
  \bibnamefont {Gamst}}, \bibinfo {author} {\bibfnamefont {K.~A.}\ \bibnamefont
  {Persson}},\ and\ \bibinfo {author} {\bibfnamefont {G.}~\bibnamefont
  {Ceder}},\ }\bibfield  {title} {\bibinfo {title} {The thermodynamic scale of
  inorganic crystalline metastability},\ }\href
  {https://advances.sciencemag.org/content/2/11/e1600225.short} {\bibfield
  {journal} {\bibinfo  {journal} {Science advances}\ }\textbf {\bibinfo
  {volume} {2}},\ \bibinfo {pages} {e1600225} (\bibinfo {year}
  {2016})}\BibitemShut {NoStop}%
\bibitem [{\citenamefont {Togo}\ and\ \citenamefont {Tanaka}(2015)}]{phonopy}%
  \BibitemOpen
  \bibfield  {author} {\bibinfo {author} {\bibfnamefont {A.}~\bibnamefont
  {Togo}}\ and\ \bibinfo {author} {\bibfnamefont {I.}~\bibnamefont {Tanaka}},\
  }\bibfield  {title} {\bibinfo {title} {First principles phonon calculations
  in materials science},\ }\href
  {https://www.sciencedirect.com/science/article/pii/S1359646215003127}
  {\bibfield  {journal} {\bibinfo  {journal} {Scr. Mater.}\ }\textbf {\bibinfo
  {volume} {108}},\ \bibinfo {pages} {1} (\bibinfo {year} {2015})}\BibitemShut
  {NoStop}%
\bibitem [{\citenamefont {Setyawan}\ and\ \citenamefont
  {Curtarolo}(2010)}]{setyawan2010high}%
  \BibitemOpen
  \bibfield  {author} {\bibinfo {author} {\bibfnamefont {W.}~\bibnamefont
  {Setyawan}}\ and\ \bibinfo {author} {\bibfnamefont {S.}~\bibnamefont
  {Curtarolo}},\ }\bibfield  {title} {\bibinfo {title} {High-throughput
  electronic band structure calculations: Challenges and tools},\ }\href
  {https://www.sciencedirect.com/science/article/pii/S0927025610002697?casa_token=zBSAY2rIt2UAAAAA:905ZMfSXRSLHZROIYJkL1OVeB2KN8WMPki_8V_o7tlAV_XeUQTnU5ox2jyiRzYf5ZwP12FLB}
  {\bibfield  {journal} {\bibinfo  {journal} {Comp. Mater. Sci.}\ }\textbf
  {\bibinfo {volume} {49}},\ \bibinfo {pages} {299} (\bibinfo {year}
  {2010})}\BibitemShut {NoStop}%
\bibitem [{\citenamefont {Chaput}\ \emph {et~al.}(2011)\citenamefont {Chaput},
  \citenamefont {Togo}, \citenamefont {Tanaka},\ and\ \citenamefont
  {Hug}}]{chaput2011phonon}%
  \BibitemOpen
  \bibfield  {author} {\bibinfo {author} {\bibfnamefont {L.}~\bibnamefont
  {Chaput}}, \bibinfo {author} {\bibfnamefont {A.}~\bibnamefont {Togo}},
  \bibinfo {author} {\bibfnamefont {I.}~\bibnamefont {Tanaka}},\ and\ \bibinfo
  {author} {\bibfnamefont {G.}~\bibnamefont {Hug}},\ }\bibfield  {title}
  {\bibinfo {title} {Phonon-phonon interactions in transition metals},\ }\href
  {https://journals.aps.org/prb/abstract/10.1103/PhysRevB.84.094302} {\bibfield
   {journal} {\bibinfo  {journal} {Phys. Rev. B}\ }\textbf {\bibinfo {volume}
  {84}},\ \bibinfo {pages} {094302} (\bibinfo {year} {2011})}\BibitemShut
  {NoStop}%
\bibitem [{\citenamefont {Togo}\ \emph {et~al.}(2015)\citenamefont {Togo},
  \citenamefont {Chaput},\ and\ \citenamefont
  {Tanaka}}]{togo2015distributions}%
  \BibitemOpen
  \bibfield  {author} {\bibinfo {author} {\bibfnamefont {A.}~\bibnamefont
  {Togo}}, \bibinfo {author} {\bibfnamefont {L.}~\bibnamefont {Chaput}},\ and\
  \bibinfo {author} {\bibfnamefont {I.}~\bibnamefont {Tanaka}},\ }\bibfield
  {title} {\bibinfo {title} {Distributions of phonon lifetimes in brillouin
  zones},\ }\href
  {https://journals.aps.org/prb/abstract/10.1103/PhysRevB.91.094306} {\bibfield
   {journal} {\bibinfo  {journal} {Phys. Rev. B}\ }\textbf {\bibinfo {volume}
  {91}},\ \bibinfo {pages} {094306} (\bibinfo {year} {2015})}\BibitemShut
  {NoStop}%
\bibitem [{\citenamefont {Zhou}\ \emph {et~al.}(2014)\citenamefont {Zhou},
  \citenamefont {Nielson}, \citenamefont {Xia},\ and\ \citenamefont
  {Ozoli{\c{n}}{\v{s}}}}]{zhou2014lattice}%
  \BibitemOpen
  \bibfield  {author} {\bibinfo {author} {\bibfnamefont {F.}~\bibnamefont
  {Zhou}}, \bibinfo {author} {\bibfnamefont {W.}~\bibnamefont {Nielson}},
  \bibinfo {author} {\bibfnamefont {Y.}~\bibnamefont {Xia}},\ and\ \bibinfo
  {author} {\bibfnamefont {V.}~\bibnamefont {Ozoli{\c{n}}{\v{s}}}},\ }\bibfield
   {title} {\bibinfo {title} {Lattice anharmonicity and thermal conductivity
  from compressive sensing of first-principles calculations},\ }\href
  {https://journals.aps.org/prl/abstract/10.1103/PhysRevLett.113.185501}
  {\bibfield  {journal} {\bibinfo  {journal} {Phys. Rev. Lett.}\ }\textbf
  {\bibinfo {volume} {113}},\ \bibinfo {pages} {185501} (\bibinfo {year}
  {2014})}\BibitemShut {NoStop}%
\bibitem [{\citenamefont {Zhou}\ \emph {et~al.}(2019)\citenamefont {Zhou},
  \citenamefont {Nielson}, \citenamefont {Xia},\ and\ \citenamefont
  {Ozoli{\c{n}}{\v{s}}}}]{zhou2019compressive1}%
  \BibitemOpen
  \bibfield  {author} {\bibinfo {author} {\bibfnamefont {F.}~\bibnamefont
  {Zhou}}, \bibinfo {author} {\bibfnamefont {W.}~\bibnamefont {Nielson}},
  \bibinfo {author} {\bibfnamefont {Y.}~\bibnamefont {Xia}},\ and\ \bibinfo
  {author} {\bibfnamefont {V.}~\bibnamefont {Ozoli{\c{n}}{\v{s}}}},\ }\bibfield
   {title} {\bibinfo {title} {Compressive sensing lattice dynamics. i. general
  formalism},\ }\href
  {https://journals.aps.org/prb/abstract/10.1103/PhysRevB.100.184308}
  {\bibfield  {journal} {\bibinfo  {journal} {Physical Review B}\ }\textbf
  {\bibinfo {volume} {100}},\ \bibinfo {pages} {184308} (\bibinfo {year}
  {2019})}\BibitemShut {NoStop}%
\bibitem [{\citenamefont {Li}\ \emph {et~al.}(2014)\citenamefont {Li},
  \citenamefont {Carrete}, \citenamefont {Katcho},\ and\ \citenamefont
  {Mingo}}]{li2014shengbte}%
  \BibitemOpen
  \bibfield  {author} {\bibinfo {author} {\bibfnamefont {W.}~\bibnamefont
  {Li}}, \bibinfo {author} {\bibfnamefont {J.}~\bibnamefont {Carrete}},
  \bibinfo {author} {\bibfnamefont {N.~A.}\ \bibnamefont {Katcho}},\ and\
  \bibinfo {author} {\bibfnamefont {N.}~\bibnamefont {Mingo}},\ }\bibfield
  {title} {\bibinfo {title} {Shengbte: A solver of the boltzmann transport
  equation for phonons},\ }\href
  {https://www.sciencedirect.com/science/article/pii/S0010465514000484?casa_token=uE9iJHa1d7gAAAAA:rLtADG7jvuwHYPjyHO-mmjqDGiIIL2Eq6ZWqJsVbE1n5fXOcyAams7aoxp0cem50ROB3EEyo}
  {\bibfield  {journal} {\bibinfo  {journal} {Comp. Phys. Comm.}\ }\textbf
  {\bibinfo {volume} {185}},\ \bibinfo {pages} {1747} (\bibinfo {year}
  {2014})}\BibitemShut {NoStop}%
\bibitem [{\citenamefont {Li}\ \emph {et~al.}(2015)\citenamefont {Li},
  \citenamefont {Hong}, \citenamefont {May}, \citenamefont {Bansal},
  \citenamefont {Chi}, \citenamefont {Hong}, \citenamefont {Ehlers},\ and\
  \citenamefont {Delaire}}]{li2015orbitally}%
  \BibitemOpen
  \bibfield  {author} {\bibinfo {author} {\bibfnamefont {C.~W.}\ \bibnamefont
  {Li}}, \bibinfo {author} {\bibfnamefont {J.}~\bibnamefont {Hong}}, \bibinfo
  {author} {\bibfnamefont {A.~F.}\ \bibnamefont {May}}, \bibinfo {author}
  {\bibfnamefont {D.}~\bibnamefont {Bansal}}, \bibinfo {author} {\bibfnamefont
  {S.}~\bibnamefont {Chi}}, \bibinfo {author} {\bibfnamefont {T.}~\bibnamefont
  {Hong}}, \bibinfo {author} {\bibfnamefont {G.}~\bibnamefont {Ehlers}},\ and\
  \bibinfo {author} {\bibfnamefont {O.}~\bibnamefont {Delaire}},\ }\bibfield
  {title} {\bibinfo {title} {Orbitally driven giant phonon anharmonicity in
  snse},\ }\href {https://www.nature.com/articles/nphys3492} {\bibfield
  {journal} {\bibinfo  {journal} {Nat. Phys.}\ }\textbf {\bibinfo {volume}
  {11}},\ \bibinfo {pages} {1063} (\bibinfo {year} {2015})}\BibitemShut
  {NoStop}%
\end{thebibliography}


\clearpage
\newpage

\end{document}